\documentclass[12pt,a4paper,twoside,english]{article}
\usepackage{times}
\usepackage[T1]{fontenc}
\usepackage[latin1]{inputenc}
\usepackage{subfigure}
\usepackage{graphicx}
\usepackage{setspace}
\usepackage{amssymb,amsmath }

\makeatletter

\usepackage{subfigure}



\usepackage{babel}
\makeatother
\begin{document}
\author{Aharon Brodutch}
\title{Weak measurements of non local variables}
\date{MSc thesis submitted May 2008}
\maketitle
\begin{center}
Thesis supervisor:Lev Vaidman
\par\end{center}

\begin{center}
School of Physics and Astronomy
\par\end{center}

\begin{center}
Raymond and Beverly Sackler Faculty of Exact
\par\end{center}

\begin{center}
Tel-Aviv University, Tel-Aviv 69978, Israel
\par\end{center}

\begin{center}
cap.fwiffo@gmail.com
\par\end{center}

This is an updated version of my MSc thesis submitted a few years ago. As I  was recently reading through I found a number of annoying  typos and decided to put a slightly modified version on arXiv. In the yeas since this thesis was submitted some progress was made regarding the measurement of non-local weak variables, in particular \cite{Kedem2010}. Some of the results in this thesis were published in \cite{Brodutch2009}.
\newpage
\begin{abstract}
Methods for measuring the weak value of non local variables are investigated.
We analyze local (indirect) measurement methods for obtaining the
weak values. We also describe some new (direct) methods (\textit{Non
local weak measurements}) for measuring the weak values of some non
local variables.

Non local variables are variables of a composite system with parts
in two or more remote locations. An example of a non local variable
is the sum of the spin components of a spin singlet state with one
spin 1/2 particle on earth and the other on the moon. Non local variables
cannot always be measured in a direct way because such measurements
can sometimes contradict relativistic causality. Those non local variables
that can be measured require the use of a specific, entangled measuring
device for the (non local) measurement .

\textit{Weak measurements} are best described as standard measurements
with a weakened coupling. Unlike the standard (strong) measurement
whose result is one of the eigenvalues, the result of weak measurements
is the \textit{weak value} , a complex number. Weak measurements have
an inherently large uncertainty, and must therefore be made on a large
ensemble of identical systems. The weak values under investigation
here are those of a non local sum and a non local product of two observables. 

For the sum of two observables we describe a method for measuring
the weak value in a direct way. We also describe an indirect method
for obtaining the weak value using the relation between the sum of
the local weak values and the weak value of the non local sum. This
indirect method requires two local measuring devices, one at each
location. The indirect (local) method is compared to the direct (non
local) method and is shown to be inefficient.

For the product of two observables, there is no direct method for
measuring the non local weak values. In some special cases the methods
we found for the measurement of a sum can be used to measure the product.
Using these methods to measure the product requires some prior knowledge
of the system to be measured.

We give a critical analysis of a recent method for measuring the product
(called {}``\textit{joint weak values}''). This method for obtaining
the weak value of the product using local weak measurements is shown
to be even more inefficient than the local method described for the
sum. We compare this method with a (non physical) direct method and
show that it requires much larger resources.

\newpage{}

\tableofcontents{}
\end{abstract}
\setlength{\topmargin}{-0.5in} \setlength{\textwidth}{6.5in} 

\section{Introduction}

Quantum measurement theory gives us the tools required for investigating
the limits of measurements under the quantum regime. The limits on
{}``What can be measured?'' and {}``with what precision?'' , have
been under investigation from the first days of quantum theory and
are still being discussed today. \cite{Landau,InstanNL,QuantumMeasurmentbook}.
One of the fundamental questions relating to our understanding of
the quantum world , the implementation of quantum computers and to
quantum information theory is that of the instantaneous measurement%
\footnote{Measurement here and throughout refers to an instantaneous measurement%
} of non local variables.

Non local variables are variables of a composite system with parts
in two or more remote locations. An example is the sum of the spin
components of a singlet state $\frac{|\uparrow\downarrow\rangle_{AB}-|\downarrow\uparrow\rangle_{AB}}{\sqrt{2}}$
where particle A (belonging to Alice) is on the moon while particle
B (belonging to Bob) is here on earth. Such variables were thought,
at first, to be unmeasurable (in an instantaneous non-demolition measurement)\cite{Landau}
since they seem to contradict relativistic causality. However it has
been shown by Aharonov Albert and Vaidman that some non local variables
can be measured instantaneously. \cite{InstanNL,AAVnonlocal,AharonovAlbert}.
As will be shown later, such measurements require a very specific
preparation of a non local measuring device (i.e an entangled system).
We will look at two types of non local variables . The sum of two
observables (eg.$\sigma_{z}^{A}+\sigma_{z}^{B}$) and the product
of two observables (eg.$\sigma_{z}^{A}\sigma_{z}^{B}$). 

There are methods for making non demolition measurements of the sum,
using an entangled measuring device prepared in a very specific way.
These methods can be used for making weak measurements if the coupling
is weakened. Another non local variable which can be measured in an
instantaneous non demolition measurement is the modular sum (eg.$(\sigma_{z}^{A}+\sigma_{z}^{B})\; mod\;4$).
This measurement requires the use of an entangled measuring device
with a very strict periodic nature which cannot be used for pointing
at weak values.

The measurement of a product in an instantaneous non demolition measurement
can sometimes allow us to send superluminal signals and thus contradict
relativistic causality. But the product can be measured in a demolition
measurement, i.e one where the final state of the system, after measurement,
is not the eigenstate corresponding to the result \cite{InstanNL}.

All measurements require some coupling between a measuring device
and the system to be measured. Van Neumann measurements\cite{vonNeumann}
are usually used to describe the measurement of a variable $A$ in
the following way: 

The measuring device is described by a pointer variable $Q$ and its
conjugate momentum $P$. It is coupled to the system via a coupling
Hamiltonian. \begin{equation}
H_{M}=g(t)PA\label{Hm}\end{equation}
Where $g(t)$ is large for a very short time around the measurement. 

When this coupling is very weak we have a new type of measurement
called a \textit{weak measurement}\cite{spin100,timeinQM}. Weak measurements
were shown to be of special interest when discussing pre and post
selected systems, systems with a definite past and a definite future
as described by a {}``Two state vector'' (TSV) \cite{ABL,timeinQM}.
When measuring these pre and post selected systems using weak measurements,
the result is something new called the \emph{Weak Value} defined as:

\begin{equation}
(A_{w})\equiv\frac{\left\langle \Phi\right|A\left|\Psi\right\rangle }{\left\langle \Phi|\Psi\right\rangle }\label{aweak}\end{equation}
This is the weak value of the observable $A$ belonging to a system
pre selected as $|\Psi\rangle$and post selected as $|\Phi\rangle$. 

One of the unique properties of weak values is that they can be far
outside the range of allowed eigenvalues. The weak measurement of
the spin of an electron can give the result 100. This strange and
surprising effect occurs when the pre and post selection are almost
orthogonal. The pointer variable of the measuring device is shifted
by this weak value as long as the coupling is weak. We can think of
the measuring device as being effectively coupled to the weak value.

We can now begin our discussion of non local weak measurements and
the measurement of non local weak values. Values of non local observables
can essentially be measured in two ways:

1) A direct measurement of the non local variable (a non local measurement)

2) An indirect measurement of the non local variable using local measurements
only. (often more than one measurement, requiring an ensemble of identically
prepared systems). 

The second method is often very disturbing, since the measurement
results in a projection to local product states. Seemingly weak measurements
are an ideal way to overcome this problem since they are inherently
non disturbing. On the other hand it seems that weak measurements
may not be effected by non local correlations for the same reason.
Nevertheless a recent method for measuring the weak value of a non
local product $(AB)_{w}$ (of two observables $A$ and $B$) makes
use of the correlations between the local measuring devices \cite{ReschSteinberg}.
This method called {}``\textit{Joint weak values}'' uses second
order effects to show that it is possible to calculate the weak value
by joining the results of two local weak measurements. Actually rather
then look at the pointer of each measuring device ($Q_{A}\;;\; Q_{B}$).
we must look at correlations between the two measuring devices $(Q_{A}Q_{B})$.
If both measuring devices are prepared in a certain way we get the
final result.\begin{equation}
Re(AB)_{w}=2\langle Q_{A}Q_{B}\rangle-Re(A_{w}^{*}B_{w})\end{equation}
As can be seen from the formula above, this is not a direct measurement
of the variable $(AB)_{w}$. In what follows we show that there is
still a major difference between such local weak measurements and
a theoretical (non-physical) non local weak measurement. We can of
course still ask ourselves if non local weak values can be measured
using a true non local weak measurement method. The problem with constructing
such non local weak measurements is that they have to follow the restrictions
of both non local measurements and weak measurements. In some cases
as in that of the modular sum these restrictions seem to be in direct
conflict with each other. So can we measure non local weak values?
can we do so using local weak measurements ? And is there a direct
method for obtaining the non local weak value? These are the questions
we want to answer.

In the next three sections we introduce the basic concepts required
to answer the above questions: non local measurements and non local
variables, the two state vector formalism, and weak measurements.
Throughout we assume that a measuring device ($Q,P$) can be coupled
to a physical system via a local operator $A$ using the coupling
Hamiltonian $H_{M}=g(t)PA$. Under this assumption we can measure
all local variables (weak or strong) directly. Quantum mechanics teaches
us that all measurements affect the system, we will discuss these
effects for both weak and strong measurements. When speaking of weak
measurements we define a new variable called the weak uncertainty
(the uncertainty in the weak value). We also describe an interesting
property of weak measurements. Weak measurements can be made on an
ensemble of random states, as long as the weak value to be measured
is the same for all systems in the ensemble. This is a consequence
of the effective coupling between the measuring device and the weak
value.

In section \ref{sec:Non-local-Measurmements-PP} we discuss non local
measurements under the two state vector formalism and define a general
classification for an efficient and inefficient measurement using
the uncertainty. We will then go on to deal with non local weak measurements
and measurements of local weak values.

In section \ref{sec:Weak-sum} we discuss the measurement of the simplest
non local weak value, that of a sum of two observables. For this simple
case, there exists a simple relation between the local and non local
weak values. $A_{w}+B_{w}=(A+B)_{w}$. We show how this can be used
to make an indirect measurement of the weak value of the sum. We also
describe non local weak measurement methods for pointing at the weak
value . These measurement methods are then compared for the general
case and for some specific examples. It is shown that the local method
is an inefficient measurement method.

In section\ref{sec:Joint-product} we discuss the method of joint
weak measurements introduced by Resch and Steinberg \cite{ReschSteinberg,Resch2}
and later refined by Resch and Lundeen \cite{ReschLundeen}. We analyze
this method and compare it to a non-physical non local method. It
is shown that such a measurement is again inefficient in all cases.
This method requires such large resources that it seems to be very
unpractical in most interesting physical situations. We also look
at sequential weak measurements \cite{squential}, and sequential
weak values. The sequential weak values are the weak values of the
product of two operators at different times $(B_{t_{2}}A_{t_{1}})_{w}=\frac{\langle\Phi|B{}_{t_{2}}VA_{t_{1}}|\Psi\rangle}{\langle\Phi|\Psi\rangle}$
where $V$ is the unitary evolution between the times $t_{1}$and
$t_{2}$. These sequential weak values are of some interest when looking
at systems where there is no meaning for a sequential {}``strong
value''. An example of this is a double interferometer experiment
where one measurement at time $t_{1}$effects the outcome of a subsequent
measurement at time $t_{2}$. Since there is no direct way of measuring
the sequential weak values, their status as {}``elements of reality''
is questioned. Finally we show that the method of joint weak measurements
cannot be used to measure the weak value of a random ensemble of system
with the same weak value.

In the last section (\ref{sec:The-meaning-of}) we discuss the meaning
of weak values. We show that in the standard weak measurement process
the measuring device is effectively coupled to the weak value. We
discuss other measurement schemes for obtaining the weak value and
show that for such measurements the weak value has a lesser status
since it is just the result of a bra-ket calculation. \newpage{}

\section{Non local measurements and non local variables\label{sec:Non-local-measurements}.}

Non local variables are unique to the quantum world. The fact that
such variables can be measured directly even in theory is not trivial
since such measurements can often contradict relativistic causality
(by allowing us to send a superluminal signal). Still some non local
variables can be measured directly as will be shown in this section.
But before we begin our description of non local measurements, let
us first review some of the principles of {}``ideal'' local measurement.

Ideal (non demolition) measurement as described by Von Neumann\cite{vonNeumann}
(also called a Von Neumann measurement) consists of a coupling interaction
between a measuring device with a pointer variable $Q$ and its conjugate
momentum $P$ and a system $|\Psi\rangle$ via a coupling, (Von Neumann)
Hamiltonian.\begin{equation}
H_{M}=g(t)PA\label{Hm}\end{equation}

where $A$ is the observable we want measured and $g(t)$is non zero
only for a short time about the measurement. For such a measurement
to be ideal, the measuring device and the system should end up (after
the interaction) in an entangled state. So that eigenstates of the
same eigenvalue will correspond to orthogonal states of the measuring
device and the amplitudes corresponding to each result remain the
same. For a state described by\begin{equation}
|\Psi\rangle_{i}=\sum\alpha_{i}|A=a_{i}\rangle\label{Psi}\end{equation}
and a measuring device $|0\rangle_{MD}$we should end up with \begin{equation}
|\Psi MD\rangle_{f}=\sum\alpha_{i}|A=a_{i}\rangle|a_{i}\rangle_{MD}\label{sysfinal}\end{equation}
where all $|a_{i}\rangle_{MD}$are orthogonal. The requirements for
such measurements are that the Hamiltonian (\ref{Hm}) can be implemented
and that the measuring device is initially in a state that is very
narrow in $Q$ compared to the interaction strength. Assuming all
such Hamiltonians can be implemented we can make all ideal measurements
with the following properties \cite{timeinQM}.

1. All systems initially in an eigenstate of $A$ will remain unchanged
by the measurement.

2. All systems not in an eigenstate of $A$ will end up (after the
measurement) in an eigenstate of $\hat{A}$ with and eigenvalue corresponding
to the result of the measurement.

Now let us assume that such measurements could be made for non local
observables such as $S_{z}^{A}S_{z}^{B}$ (the product of two spin
1 operators shared by two people, Alice and Bob) and show how we can
send a superluminal signal from Bob to Alice thus endangering relativist
causality. Before time $t=0$ we prepare the two particles in the
initial state \begin{equation}
|\Psi\rangle_{in}=\frac{1}{\sqrt{2}}(|-1\rangle_{A}+|0\rangle_{A})~|0\rangle_{B}\label{insta}\end{equation}
 We assume that at time $t=0$ somebody performs an ideal measurement
of $S_{z}^{A}S_{z}^{B}$. Immediately after $t=0$ Alice performs
a projective local measurement of the state $\frac{1}{\sqrt{2}}(|-1\rangle_{A}+|0\rangle_{A})$.
Bob, who has access to particle $B$, can send a superluminal signal
to Alice in the following way. Just before $t=0$ he decides to change
the state of his spin to $|1\rangle_{B}$ or to leave it as it is
$|0\rangle_{B}$. If he decides to do nothing, then non local measurement
of $S_{z}^{A}S_{z}^{B}$ will not change the state of the particles
because state (\ref{insta}) is an eigenstate of $S_{z}^{A}S_{z}^{B}$.
Therefore, Alice, in her local projective measurement will find the
state $\frac{1}{\sqrt{2}}(|-1\rangle_{A}+|0\rangle_{A})$ with certainty.
However, if Bob decides to change the state of his spin to $|1\rangle_{B}$
the initial state before the non local measurement at $t=0$ will
be changed to \begin{equation}
|\Psi'\rangle_{in}=\frac{1}{\sqrt{2}}(|-1\rangle_{A}+|0\rangle_{A})|1\rangle_{B}\label{insta1}\end{equation}
 It is not an eigenstate of $\sigma_{zA}\sigma_{zB}$ and thus after
the measurement the system will end up in a mixed state of $|-1\rangle_{A}|1\rangle_{B}$
or $|0\rangle_{A}|1\rangle_{B}$. In both cases the probability to
obtain a positive outcome in Alice's projective measurement on the
state $\frac{1}{\sqrt{2}}(|-1\rangle_{A}+|0\rangle_{A})$ is just
half. Instantaneous change of probability of a measurement performed
by Alice breaks causality, therefore instantaneous measurement of
$S_{z}^{A}S_{z}^{B}$ is impossible.

So can we still measure non local variables? Let us first define non
local variables as variables of a composite system with parts in two
or more remote locations. For example the sum,product or modular sum
of two spins. Such variables could always be measured in a non ideal
measurement such as a demolition measurement \cite{InstanNL}. This
method does not contradict causality, since such methods do not project
the system into the eigenstate corresponding to the measurement result.

For some non local variables such as that of a sum and a modular sum,
ideal non local measurements do not contradict causality\cite{AAVnonlocal}
and can be implemented using a non local (entangled) measuring device.
The main problem (other then braking causality) of non local measurements
is that no non local interactions exist in nature. If $O_{AB}$is
a non local operator we cannot create the measurement Hamiltonian
of the form (\ref{Hm}) $H_{M}=g(t)PO_{AB}$ instead we can only use
a Hamiltonian consisting of two local coupling terms.\begin{equation}
H_{m}^{AB}=g_{A}(t)P_{A}A+g_{B}(t)P_{B}B\label{Hnl}\end{equation}
 We assume $O_{AB}=O_{AB}(A,B)$ can be written as combination of
the local operators $A$ and $B$.( we are mainly interested in the
sum $O_{AB}=A+B$ the product $O_{AB}=AB$ and the modular sum $O_{AB}=(A+B)mod\; N$
where N is some positive number). It is easy to see that such an interaction
Hamiltonian requires two measuring devices $(Q_{A,}Q_{B})$, setting
these measuring devices in the most naive way (i.e. two local Gaussian)
will in turn give us two local measurements. For a sum and a modular
sum we can set up the two measuring devices in the following ways
in order to make the measurement non-local.

For the sum we have the measuring device initially in a state where
$Q_{A}+Q_{B}=0$ and $P_{A}-P_{B}=0$ in this case the sum after the
measurement will register on $Q_{A}+Q_{B}$. The modular sum is trickier
but can be achieved if we place the measuring device in a initial
state that has peaks at $(Q_{A}+Q_{B})mod\; N=0$ so that after the
measurement it will be impossible to distinguish between the different
peaks%
\footnote{A detailed description of such measurements is given in \cite{AAVnonlocal}%
}. Another possible method is using a measuring device with a built
in cyclic nature. An example is two spins in a maximally entangled
symmetric state as described below.

The most general state for two spin 1/2 particles can be described
as \begin{equation}
|\Psi\rangle_{AB}=\alpha|\uparrow\uparrow\rangle_{AB}+\beta|\downarrow\downarrow\rangle_{AB}+\gamma|\uparrow\downarrow\rangle_{AB}+\xi|\downarrow\uparrow\rangle_{AB}\end{equation}
we would generate a $(\sigma_{Z}^{A}+\sigma_{Z}^{A})mod\;4$ measurement
by taking our measuring device in the state\begin{equation}
|MD\rangle_{CD}=\frac{1}{\sqrt{2}}(|\uparrow\uparrow\rangle_{CD}+|\downarrow\downarrow\rangle_{CD})\end{equation}
 where C and A are in the same location and D and B are in the same
location, so that the following unitary transformation is allowed
\begin{equation}
U=e^{i\frac{\pi}{4}(I^{AC}+\sigma_{z}^{A}\sigma_{x}^{C}+\sigma_{x}^{C}+I^{BD}+\sigma_{z}^{B}\sigma_{x}^{D}+\sigma_{x}^{D})}\end{equation}
and this would be our measurement operation. It is basically a c-not
gate on each of the pairs AC and BD so it flips the spin of the MD
only if the state being measured is $\downarrow$. After the operation
the final state would be.\begin{equation}
\frac{1}{\sqrt{2}}\left(\alpha|\uparrow\uparrow\rangle_{AB}+\beta|\downarrow\downarrow\rangle_{AB}\right)\left(|\uparrow\uparrow\rangle_{CD}+|\downarrow\downarrow\rangle_{CD}\right)+\end{equation}
\[
\frac{1}{\sqrt{2}}\left(\gamma|\uparrow\downarrow\rangle_{AB}+\xi|\uparrow\downarrow\rangle_{AB}\right)\left(|\uparrow\downarrow\rangle_{CD}+|\uparrow\downarrow\rangle_{CD}\right)\]

This is of course a good measurement of the modular sum. We must still
look at the measuring device to reach the desired result, but this
can be done locally at a later time (so Alice and bob can share the
information).

The measurement procedures described above fulfill all the requirements
of ideal measurements and can be implemented on systems consisting
of more then two particles with only minor adjustments. As mentioned
above, given a large enough ensemble we can always use non-ideal measurements
to measure the non local variables. These methods become more complicated
when we move on to the pre and post selected systems as will be shown
in section \ref{sec:Non-local-Measurmements-PP}\newpage{}

\section{The two state vector formalism (TSVF) of quantum mechanics.}

The two-state vector formalism (TSVF) \cite{timeinQM} is a time-symmetric
description of the standard quantum mechanics originated in Aharonov,
Bergmann and Lebowitz \cite{ABL}. The TSVF describes a quantum system
at a particular time by two quantum states: the usual one, evolving
forward in time, defined by the results of a complete measurement
at the earlier time, and by the quantum state evolving backward in
time, defined by the results of a complete measurement at a later
time.

According to the standard quantum formalism, an ideal (Von Neumann)
measurement at time $t$ of a non degenerate variable $A$ tests for
existence at this time of the forward evolving state $|A=a\rangle$
(it yields the outcome $A=a$ with certainty if this was the state)
and creates the state evolving towards the future: \begin{equation}
|\Psi(t')\rangle=e^{-\frac{i}{\hbar}\int_{t}^{t'}Hdt}|A=a\rangle,\;\; t'>t.\label{fest}\end{equation}
 In the TSVF this ideal measurement also tests for backward evolving
state arriving from the future $~\langle A=a|$ and creates the state
evolving towards the past: \begin{equation}
\langle\Phi(t'')|=~\langle A=a|e^{\frac{i}{\hbar}\int_{t}^{t''}Hdt},\;\; t''<t.\label{best}\end{equation}

TSVF describes all quantum systems not only by the pre selected past
or the post selected future, but by both. Using this formalism pure
states can be described by their pre selected past state$|\Psi\rangle_{i}$(defined
by a measurement to be made in the past), their post selected future
state $\langle\Phi|_{f}$ (defined by a measurement made in the future)
or by a two state vector $\langle\Phi|\;\;|\Psi\rangle$. For any
strong measurement made at time t ($t_{i}<t<t_{f}$) we can use the
ABL formula \begin{equation}
{\rm Prob}(c_{n})=\frac{{|\langle\Phi|{\bf \Pi}_{C=c_{n}}|\Psi\rangle|^{2}}}{{\sum_{j}|\langle\Phi|{\bf \Pi}_{C=c_{j}}|\Psi\rangle|^{2}}}.\label{ABL}\end{equation}
 to calculate the probability of eigenvalue $c_{n}$being the result
of the measurement of the observable $C$. ($\Pi_{C=c_{n}}$are projection
operators projecting the state onto all eigenstates of $C$ with the
eigenvalue $c_{n}$). In many cases we can use this formula to predict
the outcome of a measurement with certainty ($P(C=c_{n})=1)$ even
when both pre and post selected systems are not in an eigenstate of
$C$ (eg see sec \ref{sec:Non-local-Measurmements-PP}) .%
\footnote{Another interesting example called the three box paradox is given
in \cite{timeinQM}%
}

Using the ABL formula, we can no longer use the bra-ket notation to
express expectation values, instead we can speak of the expectation
value of the measuring device. For pre and post selected systems the
expectation value of the measuring device $\langle Q\rangle$ , is
given by\begin{equation}
\langle Q\rangle=\sum_{all\; eignvalues}Prob(C=c_{n})\times c_{n}\equiv\widetilde{\langle C\rangle}\label{exp}\end{equation}
 Just by looking at this expression we can see that our usual relations
between the expectation values of observables in pre selected systems
don't hold for pre and post selected systems. As we will see later,
this may cause a problem when trying to evaluate non-local observables
using local strong measurements. \newpage{}

\section{Weak measurements and weak values\label{sec:Weak-measurments}.}

The most interesting phenomena which can be seen in the framework
of the TSVF are related to weak measurements\cite{properties,timeinQM}.
Weak measurements are best described as Von Neumann measurements with
a weakened coupling. After the interaction, all possible states of
the measuring device relating to different eigenvalues overlap, i.e
they are no longer orthogonal. Thus they cannot be distinguished from
one another with certainty. At the limit where the states almost overlap
completely, the total state of the MD and the system can be approximated
as a product state. The system is left unchanged by the measurement
and the measuring device points at some new value called a weak value
\cite{spin100}. It is not surprising that for pre selected only systems,
this weak value is the expectation value, since it is just the mean
result of all possible outcomes. For systems that are both pre and
post selected however, the weak value is given by the complex number
\begin{equation}
(A_{w})\equiv\frac{\left\langle \Phi\right|A\left|\Psi\right\rangle }{\left\langle \Phi|\Psi\right\rangle }\label{aweak}\end{equation}
where as usual $|\Psi\rangle$ is the pre selected state, $\langle\Phi|$
is the post selected state and $A$ is the observable measured. These
weak values can be outside the allowed range of eigenvalues when the
pre and post selected states are almost orthogonal $\left\langle \Phi|\Psi\right\rangle <<1$.
This interesting phenomena is an effect of the pre and post selection
causing some interference effects in the measuring device. These weak
values have been observed experimentally \cite{Weakexp1,weakexp2}.

\subsection{The weak measurement process }

For simplicity we describe the measuring device as a Gaussian around
zero%
\footnote{Unless otherwise noted, the measuring device is always prepared with
$\Delta$being the uncertainty in $Q$.%
}.

\begin{equation}
\psi_{i}^{MD}(Q)=(\Delta^{2}\pi)^{-1/4}e^{-{{Q^{2}}/{4\Delta^{2}}}}.\label{md-in}\end{equation}
The initial state before the measurement is given by $|\Psi\rangle_{i}|\psi\rangle_{i}^{MD}$

When we turn on the coupling Hamiltonian (\ref{Hm}) the evolution
of the state is given by the unitary operator $U=e^{-PA}$ (we neglect
the rest of the Hamiltonian and choose $g(t)$ so that it gives 1
after integration). The intermediate state after the measurement is
$U|\Psi\rangle_{i}|\psi\rangle_{i}^{MD}$. After post selection the
whole system (MD and $\Psi$) is left in the state $_{f}\langle\Phi|U|\Psi\rangle_{i}|\psi\rangle_{i}^{MD}|\Phi\rangle_{f}$
. For a weak enough measurement the measuring device will end up in
the state

\begin{equation}
\psi_{fin}^{MD}(Q)\approx(\Delta^{2}\pi)^{-1/4}e^{-(Q-A_{w})^{2}/{4\Delta^{2}}}\label{md-fin}\end{equation}

so that it is now a Gaussian pointing at the weak value. We now move
on to a more detailed description:

A system selected in an initial state \begin{equation}
|\Psi\rangle=\sum_{i}\alpha_{i}|A=a_{i}\rangle\label{psi-in}\end{equation}

and coupled to a measuring device (\ref{md-in}) via a coupling Hamiltonian
(\ref{Hm}) which gives rise to the unitary evolution $U=e^{-iPA}$,
the system is later post selected in the state \begin{equation}
|\Phi\rangle=\sum_{i}\beta_{i}|A=a_{i}\rangle\label{phi-fin}\end{equation}

so that we should be left with the (unnormalized) state

\begin{flushleft}
\begin{equation}
\left|\Phi\right\rangle \left\langle \Phi\right|\hat{U}\left|md\right\rangle \left|\Psi\right\rangle =(\Delta^{2}\pi)^{-1/4}\left\{ \sum_{i}\alpha_{i}\beta_{i}^{*}e^{-\frac{(Q-a_{i})^{2}}{4\Delta^{2}}}\right\} \left|\Phi\right\rangle \label{eq:mdif}\end{equation}
Since this is a product state we can write down the wave function
of the measuring device as a pure state.\begin{equation}
\psi_{fin}^{MD}(Q)=(\Delta^{2}\pi)^{-1/4}\sum_{i}\alpha_{i}\beta_{i}+\left\{ \sum_{i}\sum_{m=1}^{\infty}\alpha_{i}\beta_{i}^{*}\left[\frac{-(Q-a_{i})^{2}}{4\Delta^{2}}\right]^{m}/m!\right\} =\end{equation}

\par\end{flushleft}

\begin{flushleft}
\[
=(\Delta^{2}\pi)^{-1/4}\langle\Phi|\Psi\rangle+\]
\[
+(\Delta^{2}\pi)^{-1/4}\left\{ \sum_{m=1}^{\infty}\sum_{i}\alpha_{i}\beta_{i}^{*}\left[\frac{-(Q^{2}+ka-2Qka)}{4\Delta^{2}}\right]^{m}/m!\right\} =\]
\[
(\Delta^{2}\pi)^{-1/4}\left\{ \left\langle \Phi|\Psi\right\rangle +\sum_{m=1}^{\infty}\sum_{i}\alpha_{i}\beta_{i}^{*}\left[\frac{-(Q^{2}+a_{i}^{2}-2Qa)}{4\Delta^{2}}\right]\right\} +\]
\[
+(\Delta^{2}\pi)^{-1/4}\left\{ \sum_{m=2}^{\infty}\sum_{i}\alpha_{i}\beta_{i}^{*}\left[\frac{-(Q^{2}+a_{i}^{2}-2Qa_{i})}{4\Delta^{2}}\right]^{m}/m!\right\} =\]

\par\end{flushleft}

\begin{flushleft}
\begin{equation}
(\Delta^{2}\pi)^{-1/4}\left\langle \Phi|\Psi\right\rangle \left\{ 1+\sum_{m=1}^{\infty}\left[\frac{-(Q-(A)_{w})^{2}+[(A^{2})_{w}-(A)_{w}^{2}]}{4\Delta^{2}}\right]\right\} +\label{eq:tylor}\end{equation}
\[
+(\Delta^{2}\pi)^{-1/4}\left\{ \sum_{m=2}^{\infty}\sum_{i}\alpha_{i}\beta_{i}\left[\frac{-(Q^{2}+a_{i}-2Qa_{i})}{4\Delta^{2}}\right]^{m}/m!\right\} \approx\]
\begin{equation}
(\Delta^{2}\pi)^{-1/4}\left\langle \Phi|\Psi\right\rangle \left[e^{-(Q-A_{w})^{2}/{4\Delta^{2}}}+\frac{[(A^{2})_{w}-(A)_{w}^{2}]}{4\Delta^{2}}\right]+O(\frac{1}{\Delta^{4}})\end{equation}

\par\end{flushleft}

For large $\Delta$ this is almost a Gaussian around the weak value.
If the measurement was not disturbing, the normalization for this
expression would be $\frac{1}{\left\langle \Phi|\Psi\right\rangle }$
. The only non unitary operation is that of the post selection. But
the measurement did slightly change the system, so this is only an
approximation of the normalization. The term\begin{equation}
\frac{[(A^{2})_{w}-(A)_{w}^{2}]}{4\Delta^{2}}\label{weak uncer lead}\end{equation}
 is the leading term which effects both the normalization and the
position of the pointer and is thus the best quantitative measure
of how weak the measurement really is. The nominator in this terms
\begin{equation}
[(A^{2})_{w}-(A)_{w}^{2}]\label{weakuncer}\end{equation}
resembles the expression for the uncertainty squared. We call the
square root of (\ref{weakuncer}) the uncertainty in the weak value.

A more accurate value of the measuring device would be the state.\begin{equation}
\psi_{fin}^{MD}(Q)\approx N\left\{ (\Delta^{2}\pi)^{-1/4}\left[e^{-(Q-A_{w})^{2}/{4\Delta^{2}}}+\frac{[(A^{2})_{w}-(A)_{w}^{2}]}{4\Delta^{2}}\right]+O(\frac{1}{\Delta^{4}})\right\} \label{state-fin}\end{equation}
Where $N$ is the normalization. As we will see later, these slight
deviations from weak measurements are vital when comparing different
types of measurements. We use them to define the {}``weakness''
of the measurement. 

Since weak values are complex, $Q$ only points at the real part of
the weak value, obtaining the complex part forces us to look at the
conjugate momentum $P$. Looking at the expectation values for $P$ and
$Q$we have.\begin{equation}
Re(A_{w})=\langle Q\rangle\label{realweak}\end{equation}
\begin{equation}
Im(A_{w})=2\Delta^{2}\langle P\rangle\label{imweak}\end{equation}
Of course for a Gaussian, these expectation values are just the peak
of the Gaussian which was shifted by the weak value from zero. Vaidman
\cite{elements} argued that this obvious shift in the pointer is
what allows us to consider weak values as {}``elements of reality''.

\begin{figure}
\subfigure[ $\sigma=0.1,\langle Q \rangle = 0.2$]{\includegraphics[width=0.2\paperwidth,keepaspectratio]{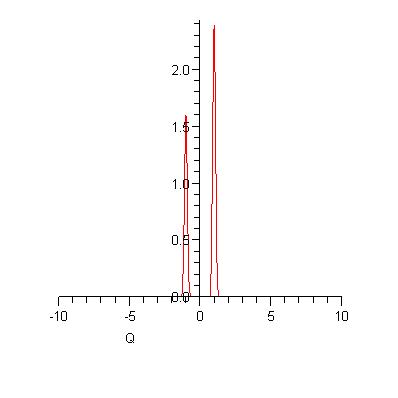}}\subfigure[ $\sigma=1 ,\langle Q \rangle = 0.5$]{\includegraphics[width=0.2\paperwidth]{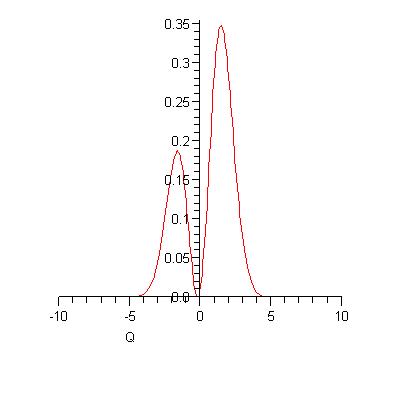}}\subfigure[ $\sigma=5 ,\langle Q \rangle=5$]{\includegraphics[width=0.2\paperwidth,keepaspectratio]{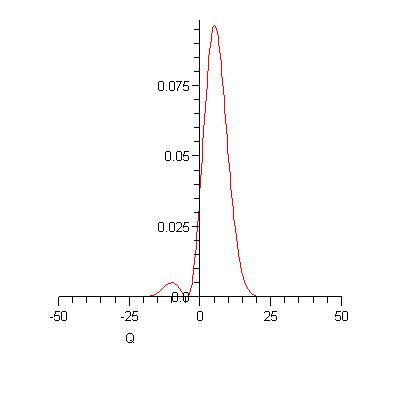}}

\subfigure[ $\sigma=25, \langle Q \rangle =9.6$]{\includegraphics[width=0.2\paperwidth]{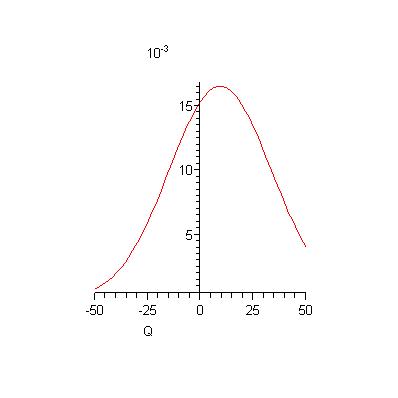}}\subfigure[ $\sigma=100,\langle Q \rangle = 9.98$]{\includegraphics[width=0.2\paperwidth,keepaspectratio]{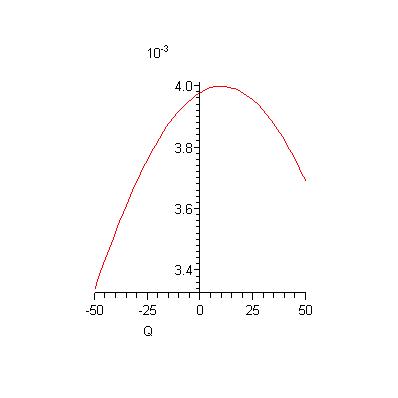}}\subfigure[ $\langle Q \rangle vs \sigma$]{\includegraphics[width=0.2\paperwidth]{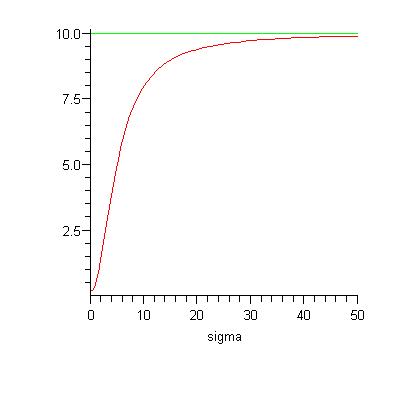}}

\caption{A measurement of $\sigma_{x}$ for the pre and post selected two
state vector $[sin\epsilon\langle\uparrow_{z}|+cos\epsilon\langle\downarrow_{z}|]\qquad|\uparrow_{z}\rangle$
with $\epsilon$ chosen so that the weak value $(\sigma_{x})_{w}=10$.
The first 5 figures (a-e) are the probability density function for
the pointer variable where sigma (the uncertainty) was varied from
0.1 in (a) corresponding to a strong measurement to 100 in (b) corresponding
to a good weak measurement as can be seen from both the graph and
the expectation value $\langle Q\rangle=9.998$. Plot (f) is the expectation
value for the measurement as a function of the uncertainty ($\sigma$)}
\end{figure}

The uncertainty in the measurement ($\Delta$) could of course be
reduced if we make a measurement of many such pre and post selected
systems. For an ensemble of $n$ such systems the uncertainty decreases
as $\frac{1}{\sqrt{n}}$. This is of course the only practical way
of making weak measurements since they inherently possess an uncertainty
large enough to make one measurement unreliable . Now, since the accuracy
of the measurement (the deviation from the weak value) depends on
$\Delta$ and the overall uncertainty depend on $\Delta$ and $n$.
We can ask ourselves a more practical question: How large an ensemble
do we need to measure the weak value with a set deviation and a set
uncertainty? We allow ourselves to be more precise and ask : How large
an ensemble do we need to measure the weak value with a deviation
of 1\% and an uncertainty of 10\% ? As will be seen in the next section,
the size of this ensemble can be used to define an efficient and an
inefficient measurement method.

The above procedure has been derived for a measuring device described
by a Gaussian, it has be shown \cite{complexweak}that any measuring
device can be used for making such weak measurements so long as the
interaction is weak enough. Thus we are reminded that a weak measurement
of an observable $A$ was defined as a standard Von Neumann measurement
made weaker. The indirect (local) methods for non local measurements
described bellow (in sections \ref{sec:Weak-sum},\ref{sec:Joint-product})
require a more lenient definition of weak measurements. There are
two requirements that must be fulfilled for the measurement to be
a weak measurement.

1. The measuring device has to point at the weak value \cite{elements}.

2. The measurement procedure must not be disturbing.

As we will see in sections \ref{sec:Weak-sum},\ref{sec:Joint-product},
these requirements are again a little strong for indirect measurements
since there is no measuring device pointing at the weak value. Still
the first requirement has some interesting consequences when looking
at ensembles of non identical systems.

\subsection{weak measurements of a random ensemble of pre and post selected systems
with the same weak value\label{sub:WMrand}.}

Usually when speaking of weak measurements, we talk of weak measurements
made on an ensemble of identically prepared (pre and post selected)
systems. An interesting change to this concept is weak measurements
made on an ensemble of random systems. These systems should be pre
and post selected in such a way that the weak value remains the same
while the pre and post selections are changed. An example of this
is an ensemble systems described by the two state vectors\begin{equation}
\langle\Phi_{j}|\;\;|\Psi_{j}\rangle=\langle\uparrow|cos(\theta_{j})+\langle\downarrow|sin(\theta_{j})\;\;\; cos(\phi_{j})|\uparrow\rangle+sin(\phi_{j})|\downarrow\rangle\label{eq:prepost-rnd}\end{equation}

with $tan(\theta_{j})tan(\phi_{j})=\frac{1}{3}$ so that the weak
value of $\sigma_{z}$ is the same for all systems.\begin{equation}
(\sigma_{z})_{w}=\frac{cos(\theta_{j})cos(\phi_{j})-sin(\theta_{j})sin(\phi_{j})}{cos(\theta)cos(\phi_{j})+sin(\theta_{j})sin(\phi_{j})}=0.5\end{equation}

Since for weak measurements the final state is a Gaussian around the
weak value, weak measurements made on all the systems will give us
the same final state for the measuring device. The result of these
weak measurements are indistinguishable from weak measurements made
on and ensemble of identically prepared pre and post selected systems.\newpage{}

\section{Non local measurements of pre and post selected systems\label{sec:Non-local-Measurmements-PP}.}

If we are only interested in the result of the measurements rather
then how much we do or do not disturb the original state it seems
that one measurement scheme is as good as another. One could argue
that such is the case for states that are pre and post selected since
we already know that we succeeded in post selection. But since non
ideal measurements project the state into a state other then the required
eigenstate, the type of measurement does effect the outcome probabilities
as can be seen by the ABL formula ( \ref{ABL}). So while joint local
measurements made on non local systems would yield the correct outcomes
for pre selected states, they don't do so for pre and post selected
systems. A simple example is a system prepared in a state \begin{equation}
|\Psi\rangle=\sqrt{\frac{1}{2+\epsilon^{2}}}(|\uparrow\rangle_{A}|\downarrow\rangle_{B}+|\downarrow\rangle_{A}|\uparrow\rangle_{B})+\epsilon|\uparrow\rangle_{A}|\uparrow\rangle_{B}\label{insta1}\end{equation}
and post selected in the state \begin{equation}
|\Phi\rangle=\sqrt{\frac{1}{2+\epsilon^{2}}}(|\uparrow\rangle_{A}|\downarrow\rangle_{B}-|\downarrow\rangle_{A}|\uparrow\rangle_{B})+\epsilon|\uparrow\rangle_{A}|\uparrow\rangle_{B}\label{finsta1}\end{equation}
with a measurement of $\sigma_{z}^{A}+\sigma_{z}^{B}$ made in the
intermediate time.

For an ideal measurement of non local variable $\sigma_{z}^{A}+\sigma_{z}^{B}$
we have the following probabilities (using \ref{ABL})

\begin{equation}
Prob(\uparrow\uparrow)_{NL}=\frac{|\langle\Phi|{\bf \Pi}_{\uparrow\uparrow}|\Psi\rangle|^{2}}{|\langle\Phi|{\bf \Pi}_{\uparrow\uparrow}|\Psi\rangle|^{2}+|\langle\Phi|{\bf \Pi}_{\downarrow\downarrow}|\Psi\rangle|^{2}+|\langle\Phi|{\bf \Pi}_{\sigma_{z}^{A}+\sigma_{z}^{B}=0}|\Psi\rangle|^{2}}=1\end{equation}
And \begin{equation}
Prob(\downarrow\downarrow)_{NL}=Prob(\sigma_{z}^{A}+\sigma_{z}^{B}=0)_{NL}=0\end{equation}

and using (\ref{exp}) \begin{equation}
\widetilde{\langle\sigma_{z}^{A}+\sigma_{z}^{B}\rangle}=2\label{non local exp1}\end{equation}

For ideal measurement of local variable $\sigma_{z}^{A}$, given that
it is the only intermediate measurement that has been performed we
have:

\begin{equation}
Prob(\uparrow)_{A}=\frac{|\langle\Phi|\Pi_{\sigma_{z}^{A}=\uparrow}|\Psi\rangle|^{2}}{|\langle\Phi|\Pi_{\sigma_{z}^{A}=\uparrow}|\Psi\rangle|^{2}+|\langle\Phi|\Pi_{\sigma_{z}^{A}=\downarrow}|\Psi\rangle|^{2}}=\frac{(\epsilon^{2}+1)^{2}}{{2+\epsilon^{4}+2\epsilon^{2}}}\label{strongsa}\end{equation}
\begin{equation}
Prob(\downarrow)_{A}=\frac{|\langle\Phi|\Pi_{\sigma_{z}^{A}=\downarrow}|\Psi\rangle|^{2}}{|\langle\Phi|\Pi_{\sigma_{z}^{A}=\uparrow}|\Psi\rangle|^{2}+|\langle\Phi|\Pi_{\sigma_{z}^{A}=\downarrow}|\Psi\rangle|^{2}}=\frac{1}{{2+\epsilon^{4}-2\epsilon^{2}}}\end{equation}
 \begin{equation}
\widetilde{\langle\sigma_{z}^{A}\rangle}=\frac{2\epsilon^{2}+\epsilon^{4}}{{2+\epsilon^{4}+2\epsilon^{2}}}\label{A LOC EXP}\end{equation}
 The expectation value of $\sigma_{z}^{B}$ is \begin{equation}
Prob(\uparrow)_{B}=\frac{|\langle\Phi|\Pi_{\sigma_{z}^{B}=\uparrow}|\Psi\rangle|^{2}}{|\langle\Phi|\Pi_{\sigma_{z}^{B}=\uparrow}|\Psi\rangle|^{2}+|\langle\Phi|\Pi_{\sigma_{z}^{B}=\downarrow}|\Psi\rangle|^{2}}=\frac{(\epsilon^{2}-1)^{2}}{{2+\epsilon^{4}-2\epsilon^{2}}}\label{strongsb}\end{equation}
\[
Prob(\downarrow)_{B}=\frac{|\langle\Phi|\Pi_{\sigma_{z}^{B}=\downarrow}|\Psi\rangle|^{2}}{|\langle\Phi|\Pi_{\sigma_{z}^{B}=\uparrow}|\Psi\rangle|^{2}+|\langle\Phi|\Pi_{\sigma_{z}^{B}=\downarrow}|\Psi\rangle|^{2}}=\frac{1}{{2+\epsilon^{4}-2\epsilon^{2}}}\]
 \begin{equation}
\widetilde{\langle\sigma_{z}^{B}\rangle}=\frac{-2\epsilon^{2}+\epsilon^{4}}{{2+\epsilon^{4}-2\epsilon^{2}}}\label{b loc exp}\end{equation}
 It is easy to see that the expectation value of the non local variable
(\ref{non local exp1}) is very different from the sum of (\ref{A LOC EXP})
and (\ref{b loc exp}). However, it is more reasonable to compare
(\ref{non local exp1}) with the sum of expectation values of the
outcomes of local measurements of $\sigma_{z}^{A}$ and $\sigma_{z}^{B}$
performed simultaneously ( a joint measurement). This corresponds
to measurement of a variable with non degenerate eigenstates $|\uparrow\rangle_{A}|\downarrow\rangle_{B},|\downarrow\rangle_{A}|\uparrow\rangle_{B},|\uparrow\rangle_{A}|\uparrow\rangle_{B},|\downarrow\rangle_{A}|\downarrow\rangle_{B}$
. In this case we have \begin{eqnarray}
\nonumber \\ & Prob(\uparrow\uparrow)_{joint}=\frac{|\langle\Phi|{\bf \Pi}_{\uparrow\uparrow}|\Psi\rangle|^{2}}{|\langle\Phi|{\bf \Pi}_{\uparrow\uparrow}|\Psi\rangle|^{2}+|\langle\Phi|{\bf \Pi}_{\uparrow\downarrow}|\Psi\rangle|^{2}+|\langle\Phi|\Pi_{\downarrow\uparrow}|\Psi\rangle|^{2}}=\frac{\epsilon^{2}}{{2+\epsilon^{4}}}\label{strongsumsa+sb}\end{eqnarray}
\[
Prob(\uparrow\downarrow)_{joint}=\frac{|\langle\Phi|{\bf \Pi}_{\uparrow\downarrow}|\Psi\rangle|^{2}}{|\langle\Phi|{\bf \Pi}_{\uparrow\uparrow}|\Psi\rangle|^{2}+|\langle\Phi|{\bf \Pi}_{\uparrow\downarrow}|\Psi\rangle|^{2}+|\langle\Phi|{\bf \Pi}_{\downarrow\uparrow}|\Psi\rangle|^{2}}=\frac{1}{{2+\epsilon^{4}}}\]
\[
Prob(\downarrow\uparrow)_{joint}=\frac{|\langle\Phi|{\bf \Pi}_{\downarrow\uparrow}|\Psi\rangle|^{2}}{|\langle\Phi|{\bf \Pi}_{\uparrow\uparrow}|\Psi\rangle|^{2}+|\langle\Phi|{\bf \Pi}_{\uparrow\downarrow}|\Psi\rangle|^{2}+|\langle\Phi|{\bf \Pi}_{\downarrow\uparrow}|\Psi\rangle|^{2}}=\frac{1}{{2+\epsilon^{4}}}\]

\begin{equation}
\widetilde{\langle\{\sigma_{z}^{A}\}+\{\sigma_{z}^{B}\}\rangle}=\frac{\epsilon^{2}}{{2+\epsilon^{4}}}\label{joint exp}\end{equation}
 The brackets $\{\}$ signify a separate measurement of each observable.
We can easily see that not only do the measurement expectation values
come out wrong, but when making local measurements we have a finite
probability to get the results $\sigma_{z}^{A}+\sigma_{z}^{B}=0$
corresponding to $\uparrow\downarrow$ and $\downarrow\uparrow$.
These have zero probability for the non local measurement. If we had
a large enough ensemble we could perform a complicated set of local
measurements and find out the information about the pre and post selected
states \cite{nonlocaltimesymmetric}. (Such measurements would often
involve some intermediate measurement used to {}``delete'' the past
or the future.) So although the first statement made in this section
may be true, there is a cost to making non ideal measurements. 

Many measurement methods require an ensemble to improve accuracy.
We can define an efficient measurement as one requiring a relatively
small ensemble to get a good accuracy (as defined by us) and an inefficient
measurement as one requiring a much larger ensemble for achieving
the same accuracy. We can also say that an efficient measurement is
one where size of the ensemble required to reach a certain accuracy
depends on the uncertainty of the variable to be measured. An inefficient
measurement would be one that depends on other parameters (such as
the uncertainty of other variables). Going back to our example we
see that the uncertainty is zero and therefore one measurement should
be sufficient. Thus an efficient measurement method should require
one measurement to arrive at the right result with a good accuracy.
A few more measurements can be made to improve the statistical significance.

For weak measurements we can use the weak uncertainty (\ref{weakuncer}
) to classify our measurement scheme as efficient or inefficient.
An efficient measurement would be one where the ensemble depends on
the weak uncertainty as in the local weak measurements defined in
section \ref{sec:Weak-measurments}. As we will see in the next section
local weak measurements of non local weak values depend on the local
weak uncertainty rather then the non local ones, making them inefficient
measurements.\newpage{}

\section{Weak measurements of a non local sum\label{sec:Weak-sum}.}

We will start with the simplest case of a non local weak measurement,
that of a sum of two observables. This case is relatively simple since
there is a simple relation between the local and non local weak values.\begin{equation}
(A+B)_{w}=A_{w}+B_{w}\label{weaksum}\end{equation}
 It is just straightforward from (\ref{aweak}). Using this formula,
we have a simple method for measuring the non local weak value just
by measuring the local weak values and adding the results. But as
we will see this is an inefficient measurement (see sec \ref{sec:Non-local-Measurmements-PP}
) since the ensemble required to make a good precision measurement
does not depend only on the uncertainty in $(A+B)_{w}$. A detailed
derivation of this measurement process will give us some insight into
the difference between the local measurements and the non local ones.

We start with the general state\begin{equation}
|\Psi\rangle=\sum_{i}\alpha_{ij}|A=a_{i},B=b_{j}\rangle\label{psi-in2}\end{equation}

and coupled to a measuring device 

\begin{equation}
\psi_{AB-in}^{MD}(Q)=(\Delta^{2}\frac{\pi}{2})^{-1/2}e^{-(Q_{A}^{2}+Q_{B}^{2})/{2\Delta^{2}}}\label{md-in2}\end{equation}

via a coupling Hamiltonian which gives rise to the unitary evolution
$\hat{U}=e^{-i(P_{A}A+P_{B}B)}$ (note that the overall uncertainty
remains $\Delta$although each Gaussian is narrower by a factor of
$\sqrt{2}$). The system is later post selected in the state \begin{equation}
|\Phi\rangle=\sum_{i}\beta_{ij}|A=a_{i},B=b_{j}\rangle\label{phi-fin2}\end{equation}
After post selection we are left with the unnormalized state \begin{equation}
\left|\Phi\right\rangle \left\langle \Phi\right|\hat{U}\left|md\right\rangle \left|\Psi\right\rangle =(\frac{2}{\pi\Delta^{2}})^{1/2}\left\{ \sum_{i}\alpha_{ij}\beta_{ij}e^{-\frac{(Q_{A}-a_{i})^{2}+(Q_{B}+b_{j})^{2}}{2\Delta^{2}}}\right\} \left|\Phi\right\rangle \label{afterpost2}\end{equation}
 expanding the measuring device's wave function as a Taylor series
we get 
\begin{align}
\psi_{fin}^{MD}(Q)\approx&\\\nonumber& (\Delta^{2}\pi)^{-1/4}\left[e^{-(Q-A_{w})^{2}/{4\Delta^{2}}}+\left\langle \Phi|\Psi\right\rangle \frac{[(A^{2})_{w}-(A)_{w}^{2}]}{2\Delta^{2}}+\left\langle \Phi|\Psi\right\rangle \frac{[(B^{2})_{w}-(B)_{w}^{2}]}{2\Delta^{2}}\right]\\\nonumber&+O(\frac{1}{\Delta^{4}}) \label{state-fin2}
\end{align}
 it is not surprising that such a method depends on the local weak
uncertainty rather then the non local one. (By local uncertainty we
mean the uncertainty in local variables). 

Using the above measuring scheme and the relation $(A+B)_{w}=A_{w}+B_{w}$
we can arrive at the weak value of the sum via local weak measurements
alone. We will use the following example to show how this method differs
from a direct (non local) measurement. 

The state is prepared initially (pre selected) in the state\begin{equation}
|\Psi\rangle_{AB}=\frac{1}{\sqrt{2\epsilon^{2}+\delta^{2}+2}}\left|\uparrow\downarrow(1+\epsilon)+\downarrow\uparrow(-1+\epsilon)+\delta\uparrow\uparrow\right\rangle \label{eq:expre}\end{equation}

and post selected in the state \begin{equation}
|\Phi\rangle_{AB}=\frac{1}{2}|\uparrow\downarrow+\downarrow\uparrow+\uparrow\uparrow+\downarrow\downarrow\rangle\label{eq:expost}\end{equation}

so that an intermediate weak measurement of $(\sigma_{z}^{A}+\sigma_{z}^{B})$would
give

$(\sigma_{z}^{A}+\sigma_{z}^{B})_{w}=\frac{\left\langle \uparrow\downarrow+\downarrow\uparrow+\uparrow\uparrow+\downarrow\downarrow\right|\sigma_{z}^{A}+\sigma_{z}^{B}\left|\uparrow\downarrow(1+\epsilon)+\downarrow\uparrow(-1+\epsilon)+\delta\uparrow\uparrow\right\rangle }{\left\langle \uparrow\downarrow+\downarrow\uparrow+\uparrow\uparrow|\uparrow\downarrow(1+\epsilon)+\downarrow\uparrow(-1+\epsilon)+\delta\uparrow\uparrow\right\rangle }=\frac{2\delta}{2\epsilon+\delta}$ 

Using this example we can now compare two measurement methods. 

1. A joint local measurement (involving two local measuring devices)
as described above.

2. A direct measurements (a non local measurement).

\begin{figure}
\subfigure[$\Delta=0.1 ; \; \langle Q \rangle = 1.1$]{\includegraphics[scale=0.3]{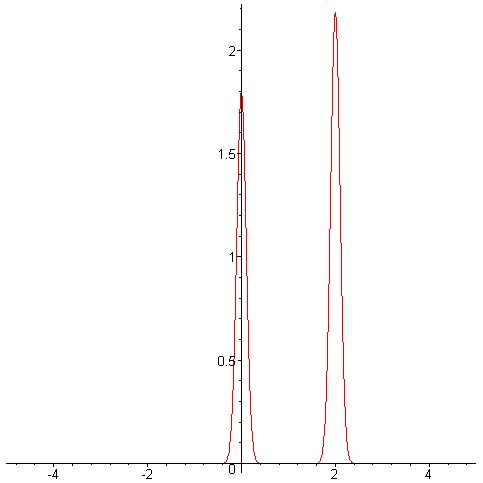}}\subfigure[$\Delta=0.1 ; \; \langle Q_A \rangle=-0.10$]{\includegraphics[scale=0.3]{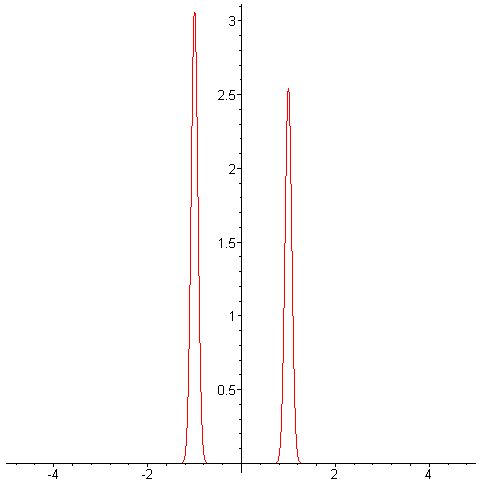}}\subfigure[$\Delta=0.1 ; \; \langle Q_B \rangle =+0.10$]{\includegraphics[scale=0.3]{\string"loc,d=11e=-05,sigma=0,1,exp=-0,1\string".jpg}}

\subfigure[$\Delta=1 ; \; \langle Q \rangle = 1.2$]{\includegraphics[scale=0.3]{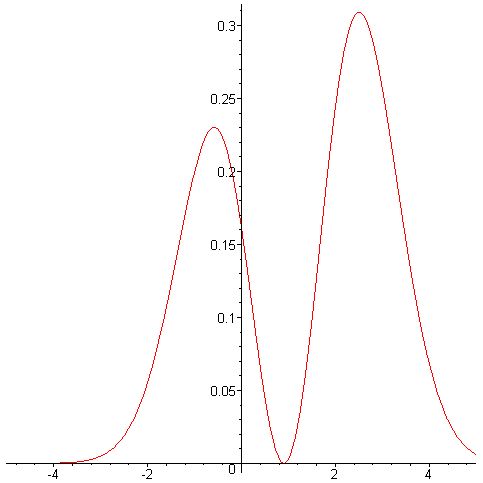}}\subfigure[$\Delta=0.1 ; \; \langle Q_A \rangle=-0.06$]{\includegraphics[scale=0.3]{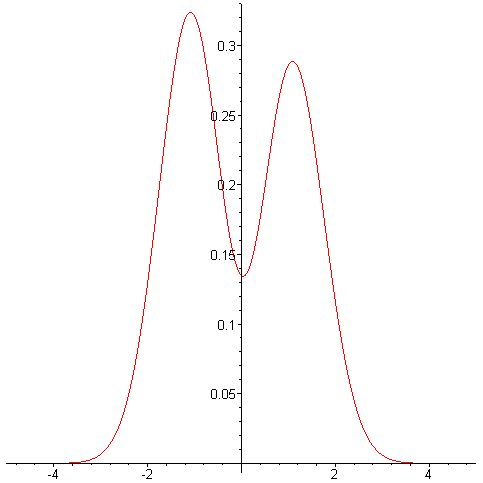}}\subfigure[$\Delta=1 ; \; \langle Q_1a \rangle =+0.07$]{\includegraphics[scale=0.3]{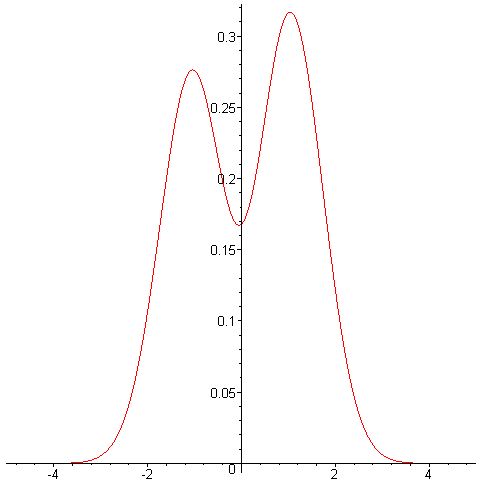}}

\subfigure[$\Delta=10; \; \langle Q \rangle = 11$]{\includegraphics[scale=0.3]{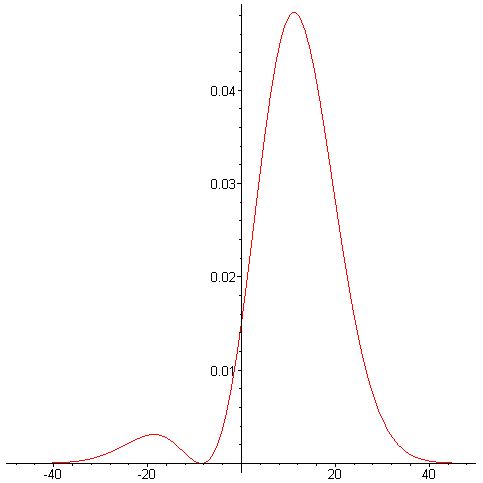}}\subfigure[$\Delta=10 ; \; \langle Q_A \rangle=+0.5$]{\includegraphics[scale=0.3]{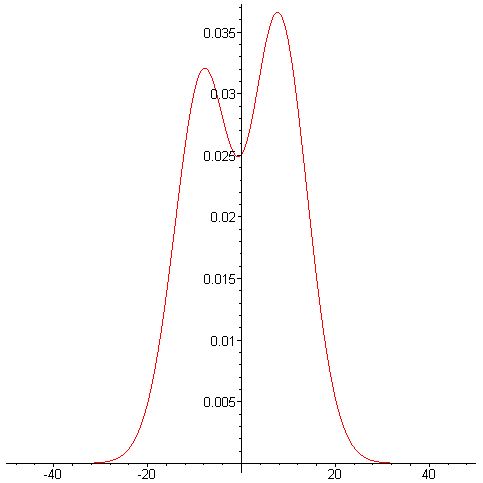}}\subfigure[$\Delta=10 ; \; \langle Q_B \rangle =-0.4$]{\includegraphics[scale=0.3]{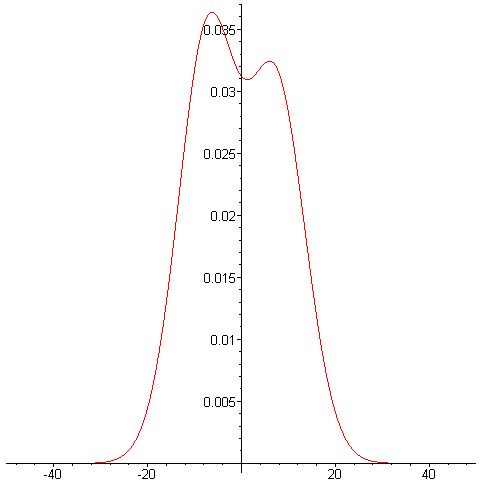}}

\caption{\label{fig:locvnloc}Comparing the local and non local methods for
reaching the weak value of a sum. $(\sigma_{z}^{A}+\sigma_{z}^{B})_{w}=\frac{\left\langle \uparrow\downarrow+\downarrow\uparrow+\uparrow\uparrow+\downarrow\downarrow\right|\sigma_{z}^{A}+\sigma_{z}^{B}\left|\uparrow\downarrow(1+\epsilon)+\downarrow\uparrow(-1+\epsilon)+\delta\uparrow\uparrow\right\rangle }{\left\langle \uparrow\downarrow+\downarrow\uparrow+\uparrow\uparrow|\uparrow\downarrow(1+\epsilon)+\downarrow\uparrow(-1+\epsilon)+\delta\uparrow\uparrow\right\rangle }=\frac{2\delta}{2\epsilon+\delta}$}

For $\delta=0.11$, $\epsilon=-0.05$ we have $(\sigma_{z}^{A}+\sigma_{z}^{B})_{w}=22$.
The local weak values are $(\sigma_{z}^{A})_{w}=211$, $(\sigma_{z}^{B})_{w}=-189$.
The probability functions for each of the measuring devices are plotted
for various values of $\Delta$(the uncertainty in measurement) with
the expectation value for the measurement given at the bottom. The
left most column (a,d,g) is the non local measuring device, the center
column (b,e,h) is the local measuring device at A, and the right column
(c,f,i) is the local measuring device at B. In this figure we look
at small values of $\Delta$. 

$\Delta=0.1$ (a,b,c) is a strong measurement. 

$\Delta=1$ (e,f,g) is a slightly weaker measurement, where we can
already see the interference effect of weak measurements. 

For $\Delta=10$ (h,i,j) we can see that the non local measuring device
(j) has an expectation value outside the range of eigenvalues. (for
$\Delta=100,100$ see fig \ref{fig:locvnloc2})
\end{figure}

\begin{figure}
\subfigure[$\Delta=100; \; \langle Q \rangle = 21.8$]{\includegraphics[scale=0.3]{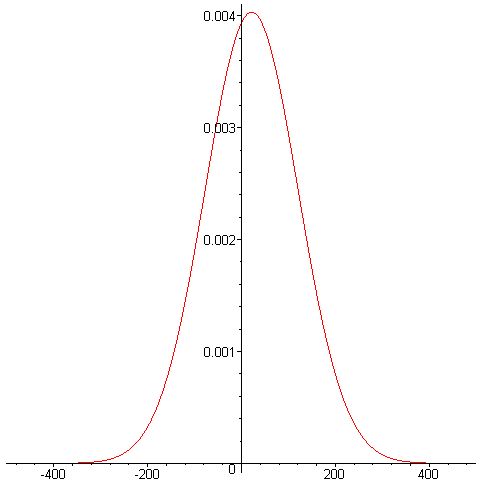}}\subfigure[$\Delta=100 ; \; \langle Q_A \rangle=42$]{\includegraphics[scale=0.3]{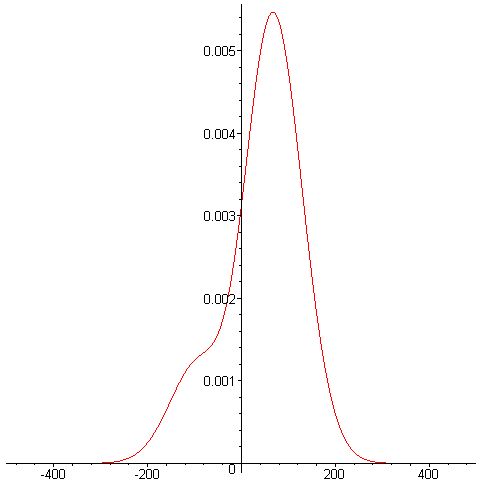}}\subfigure[$\Delta=100 ; \; \langle Q_B \rangle =-38$]{\includegraphics[scale=0.3]{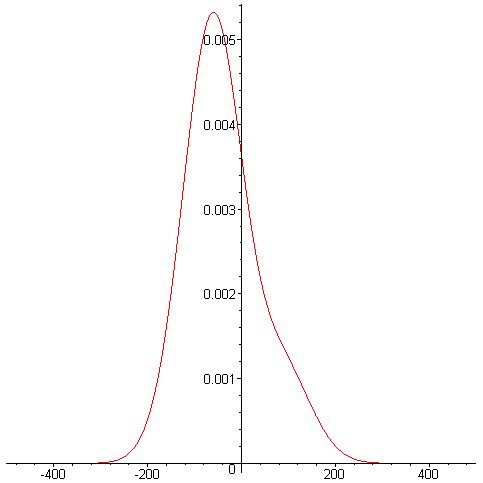}}

\subfigure[$\Delta=1000; \; \langle Q \rangle =21.998$]{\includegraphics[scale=0.3]{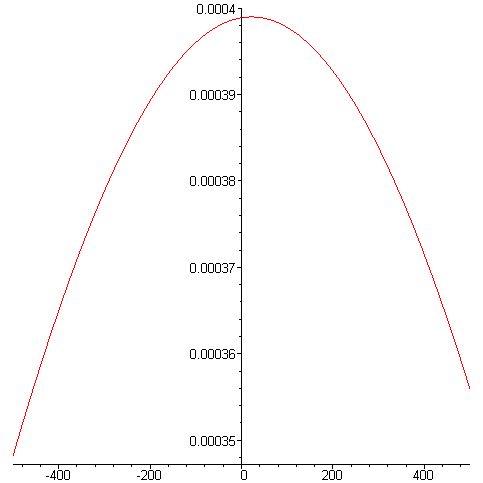}}\subfigure[$\Delta=1000 ; \; \langle Q_A \rangle=203$]{\includegraphics[scale=0.3]{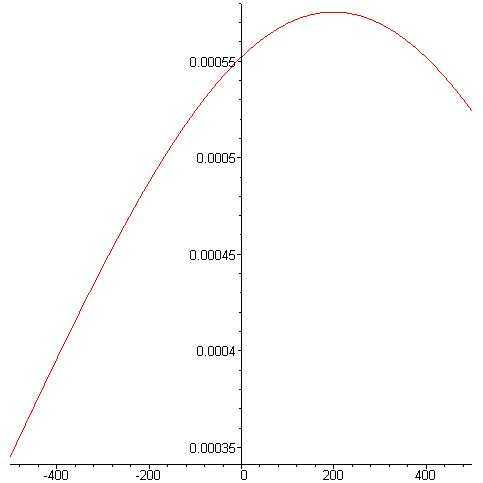}}\subfigure[$\Delta=1000; \; \langle Q_B \rangle =-181$]{\includegraphics[scale=0.3]{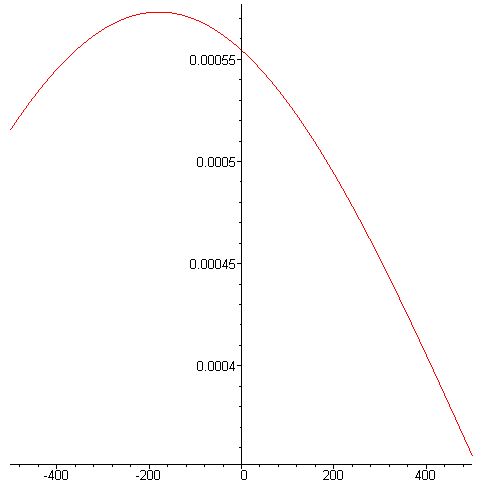}}

\caption{\label{fig:locvnloc2}Comparing the local and non local methods for
reaching the weak sum for $(\sigma_{z}^{A}+\sigma_{z}^{B})_{w}=\frac{\left\langle \uparrow\downarrow+\downarrow\uparrow+\uparrow\uparrow+\downarrow\downarrow\right|\sigma_{z}^{A}+\sigma_{z}^{B}\left|\uparrow\downarrow(1+\epsilon)+\downarrow\uparrow(-1+\epsilon)+\delta\uparrow\uparrow\right\rangle }{\left\langle \uparrow\downarrow+\downarrow\uparrow+\uparrow\uparrow|\uparrow\downarrow(1+\epsilon)+\downarrow\uparrow(-1+\epsilon)+\delta\uparrow\uparrow\right\rangle }=\frac{2\delta}{2\epsilon+\delta}$}

For $\delta=0.11$, $\epsilon=-0.05$ we have $(\sigma_{z}^{A}+\sigma_{z}^{B})_{w}=22$
while the local weak values are $(\sigma_{z}^{A})_{w}=211$, $(\sigma_{z}^{B})_{w}=-189$

The probability wave functions for each of the measuring devices are
plotted for various values of $\Delta$(the uncertainty in measurement)
with the expectation value for the measurement given at the bottom.
The left most column (a,d) is the non local measuring device, the
center column (b,e) is the local measuring device at A, and the right
column (c,f) is the local measuring device at B. In this figure we
look at relatively large values of $\Delta$. $\Delta=100$ (a,b,c)
is a good weak measurement for the non local measurement (a) while
the non local values are still off and are still not a Gaussian. For
$\Delta=1000$ (e,f,g) we have Gaussian pointing close to the weak
values for all three measurements. (for the stronger values$\Delta=0.1,1,10$
see fig \ref{fig:locvnloc})
\end{figure}

The density function of the measuring devices for the two measurement
methods (joint local method and non local method) are plotted in figs
\ref{fig:locvnloc} and \ref{fig:locvnloc2} for different values
of $\Delta$(the uncertainty in measurement) with $\epsilon=-0.05$
and $\delta=0.11.$ This is an interesting example since the local
uncertainty is much larger then the non local uncertainty. The plots
are for each measuring device, i.e the non local measuring device
in the case of the non local method and each of the local measuring
devices in the case of the local method. For low values of $\Delta$
(fig \ref{fig:locvnloc} ) we can see that both methods don't approximate
a Gaussian for the final state of the measuring device. Still it is
obvious that the non local method converges much faster to a Gaussian
then the local method, this is also true for larger values of $\Delta$
(fig \ref{fig:locvnloc2}). 

\begin{figure}
\subfigure[$\delta=0.11;\epsilon=-0.05 \; \langle Q \rangle \rightarrow 22$]{\includegraphics[scale=0.4]{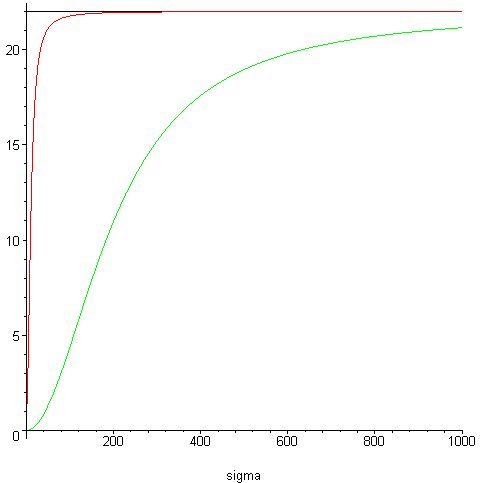}}\subfigure[$\delta=0.11;\epsilon=-0.05 \; \langle Q \rangle \rightarrow 22;\;(0-100)$]{\includegraphics[scale=0.4]{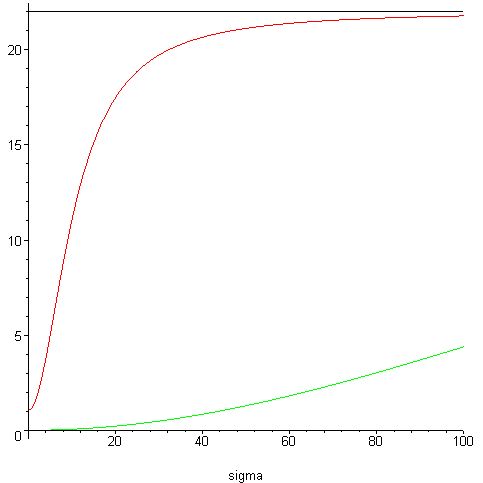}}

\caption{\label{fig:locvnlocexp}Comparing the local (green line) and non
local (red line) methods for reaching the weak value of a sum.$(\sigma_{z}^{A}+\sigma_{z}^{B})_{w}=\frac{\left\langle \uparrow\downarrow+\downarrow\uparrow+\uparrow\uparrow+\downarrow\downarrow\right|\sigma_{z}^{A}+\sigma_{z}^{B}\left|\uparrow\downarrow(1+\epsilon)+\downarrow\uparrow(-1+\epsilon)+\delta\uparrow\uparrow\right\rangle }{\left\langle \uparrow\downarrow+\downarrow\uparrow+\uparrow\uparrow|\uparrow\downarrow(1+\epsilon)+\downarrow\uparrow(-1+\epsilon)+\delta\uparrow\uparrow\right\rangle }=\frac{2\delta}{2\epsilon+\delta}$}

The plots are the expectation values at different ranges of the uncertainty
(a, $\sigma=0..1000$ b, $\sigma=0..100$). For $\delta=0.11$, $\epsilon=-0.05$
we have $(\sigma_{z}^{A}+\sigma_{z}^{B})_{w}=22$. The local weak
values are $(\sigma_{z}^{A})_{w}=211$, $(\sigma_{z}^{B})_{w}=-189$
(plots for the probability density can be seen in fig \ref{fig:locvnloc},\ref{fig:locvnloc2})

It can easily be seen that the non local method (red line) converges
much faster then the local method (green line). If we want to make
a precise measurements (deviation of 1\% and uncertainty of 10\%)
an ensemble of $n=2.2\times10^{3}$ is required for the non local
method while an ens amble of $n=8.2\times10^{5}$is required for the
non local method.
\end{figure}

\begin{figure}
\subfigure[$\delta=0.1;\epsilon=0 \; \langle Q \rangle \rightarrow 2$]{\includegraphics[scale=0.4]{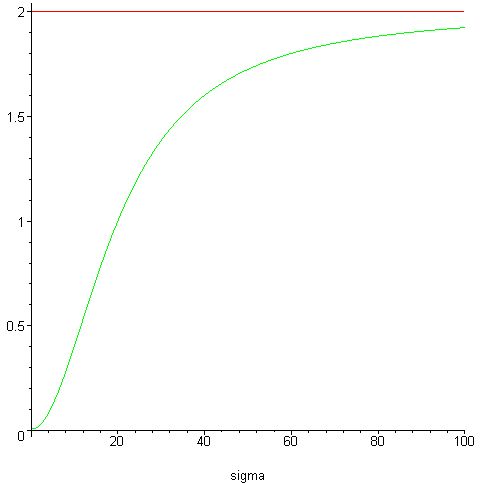}}\subfigure[$\delta=-1;\epsilon=0.6 \; \langle Q \rangle \rightarrow -10$]{\includegraphics[scale=0.4]{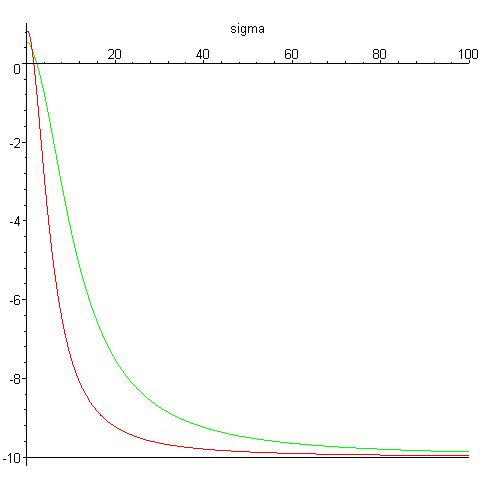}}

\subfigure[$\delta=1;\epsilon=1 \; \langle Q \rangle \rightarrow 0.67$]{\includegraphics[scale=0.4]{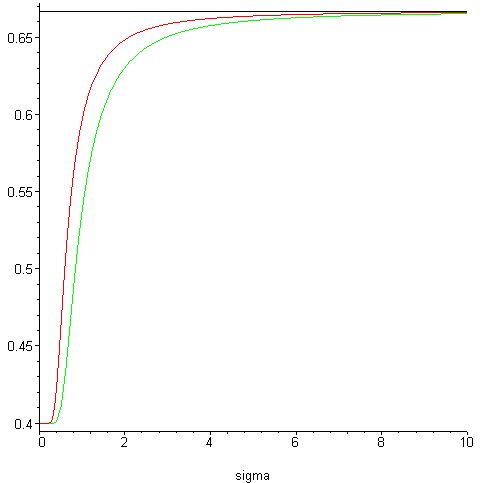}}\subfigure[$n for \epsilon=-0.05 ,\delta=0.1001..1$]{\includegraphics[scale=0.62]{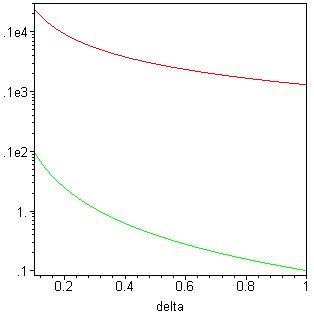}}

\caption{\label{fig:locvnlocexp2}Comparing the local (green line) and non
local (red line) methods for reaching the weak value of a sum. $(\sigma_{z}^{A}+\sigma_{z}^{B})_{w}=\frac{\left\langle \uparrow\downarrow+\downarrow\uparrow+\uparrow\uparrow+\downarrow\downarrow\right|\sigma_{z}^{A}+\sigma_{z}^{B}\left|\uparrow\downarrow(1+\epsilon)+\downarrow\uparrow(-1+\epsilon)+\delta\uparrow\uparrow\right\rangle }{\left\langle \uparrow\downarrow+\downarrow\uparrow+\uparrow\uparrow|\uparrow\downarrow(1+\epsilon)+\downarrow\uparrow(-1+\epsilon)+\delta\uparrow\uparrow\right\rangle }=\frac{2\delta}{2\epsilon+\delta}$}

The plots are the expectation values vs the uncertainty for different
values of $\epsilon$ and $\delta$ corresponding to different local
and non local weak values (non local=red line; local=green line). 

a) $\delta=0.1$, $\epsilon=0$ - $(\sigma_{z}^{A}+\sigma_{z}^{B})_{w}=2$
while the local weak values are $(\sigma_{z}^{A})_{w}=21$, $(\sigma_{z}^{B})_{w}=-19$.
This is a special case where a strong non local measurement will with
certainty give the value of 2 (as was seen in section \ref{sec:Non-local-measurements}). 

b) $\delta=-1$, $\epsilon=0.6$ - $(\sigma_{z}^{A}+\sigma_{z}^{B})=-10$
while the local weak values are $(\sigma_{z}^{A})_{w}=5$, $(\sigma_{z}^{B})_{w}=-15$.
Even though the local and non local weak values are of the same order
of magnitude, the non local method is clearly much better then the
local one.

c) $\delta=1$, $\epsilon=1$ - $(\sigma_{z}^{A}+\sigma_{z}^{B})_{w}=\frac{2}{3}$
while the local weak values are $(\sigma_{z}^{A})_{w}=1$, $(\sigma_{z}^{B})_{w}=-\frac{1}{3}$.
In this case there is a very small uncertainty in both the local and
non local weak values. There is still some advantage to the non local
measurement but it is very small.

d) A logarithmic plot of the size of the ensemble required for a precise
measurement (deviation of 1\% and uncertainty of 10\%) for $\epsilon=-0.05$
and $\delta$ Going from $0.1001$ to $1$ (at $\delta=1$ the uncertainty
in the non local measurement is very low, we are almost certainly
at the $\uparrow\uparrow$state). 
\end{figure}

Plots for the expectation values as a function of $\Delta$ are given
in fig \ref{fig:locvnlocexp}, again we see that the non local method
converges a lot faster then the local method. If we want to make a
precise measurements (deviation of 1\% and uncertainty of 10\%) we
need an ensemble of $n=2.2\times10^{3}$ systems for the non local
method while a much larger ensemble $n=8.2\times10^{5}$ is required
for the joint local method. More examples with different $\delta$
and $\epsilon$ are given in fig \ref{fig:locvnlocexp2} a,b,c. In
fig:\ref{fig:locvnlocexp2} (d) we have a plot of the ensemble required
to make a precise measurement of the same example with $\epsilon=-0.05$
and $\delta$ ranging from 0.1001 to 1.

We can see that for the general case of two measuring devices the
measurement error is at best linearly proportional to the two local
weak values. This type of measurement is an inefficient measurement
regardless of whether we can derive the desired result after the measurement
process.


 For now , the non local measurement could have been seen as just
an (unphysical) reference point since we did not yet describe a method
for making such a non local measurement. We can still use this reference
point to show that the measurements of non local values using local
measuring devices are inefficient measurements. But we can build a
good non local measuring device for measuring the non local weak sum.
We will now describe two methods for setting up the entangled measuring
device for this measurement. The first is a wide Gaussian around $Q_{A}+Q_{B}=0$,
$P_{A}-P_{B}=0$ the second is an entangled measuring device made
up of discrete overlapping Gaussian states so that again they obey
$Q_{A}+Q_{B}=0$ but each will be locally around different values.
Here the local uncertainty of the measuring device is much larger
then the non local uncertainty.

\newpage

\subsection{A Gaussian around $Q_{A}+Q_{B}=0$, $P_{A}-P_{B}=0$}

We start by defining two orthogonal variables \begin{equation}
Q^{+}=Q_{A}+Q_{B}\label{qplus}\end{equation}
\begin{equation}
Q^{-}=Q_{A}-Q_{B}\label{qminus}\end{equation}

The initial state of the measuring device is 

\begin{equation}
\psi_{AB-in}^{MD}(Q)=(\Delta\Sigma\pi)^{-1/2}e^{-(Q^{+})^{2}/{4\Delta^{2}}}e^{-(Q^{-})^{2}/{4\Sigma^{2}}}\label{Md ent}\end{equation}
We are interested in the limit $\Sigma\rightarrow\infty$. The uncertainty
in $Q^{+}$ is $\Delta$ as always. Using this measuring device and
the usual interactions $\hat{U}$ we have after pre selection\begin{equation}
|\Psi\rangle=\sum_{i}\alpha_{ij}|A=a_{i},B=b_{j}\rangle\label{psi-in2}\end{equation}
 measurement

\begin{equation}
\hat{U}=e^{-i(P_{A}A+P_{B}B)}\label{nonlocalint}\end{equation}
and post selection

\begin{equation}
|\Phi\rangle=\sum_{i}\beta_{ij}|A=a_{i},B=b_{j}\rangle\label{phi-fin2}\end{equation}

the final state (where we just traced out $Q^{-}$) \begin{equation}
\left|\Phi\right\rangle \left\langle \Phi\right|\hat{U}\left|md\right\rangle \left|\Psi\right\rangle =(\Delta^{2}\pi)^{-1/4}\left\{ \sum_{i}\alpha_{ij}\beta_{ij}e^{-\frac{(Q^{+}-a_{i}-b_{j})^{2}}{2\Delta^{2}}}\right\} \left|\Phi\right\rangle =\label{afterpost3}\end{equation}
it is very simple to see that opening this as a Taylor series would
give us a Gaussian around the non local weak value.\begin{equation}
\psi_{fin}^{MD}(Q)\approx\left\{ (\Delta^{2}\pi)^{-1/4}\left[e^{-(Q^{+}-A_{w})^{2}/{4\Delta^{2}}}+\left\langle \Phi|\Psi\right\rangle \frac{[([A+B]^{2})_{w}-(A+B)_{w}^{2}]}{2\Delta^{2}}\right]+O(\frac{1}{\Delta^{4}})\right\} \label{mdfin-ent1}\end{equation}

\subsection{A sum of Gaussians.}

Another method for making a non local measuring device is by setting
it up in the following way\[
\psi_{in}^{MD}(Q)=\frac{1}{N}\sum_{l=0}^{k}e^{-\frac{(Q_{1}+l\xi)^{2}+(Q_{2}-l\xi)^{2}}{2\Delta^{2}}}\]
 with \[
N^{2}={\displaystyle \sum_{i=0}^{m}\sum_{j=0}^{m}\pi\Delta^{2}e}xp[\frac{-\xi^{2}(i-j)^{2}}{2\Delta^{2}}]\]
 this wave function is localized around $Q_{A}+Q_{B}=0$ with the
uncertainty being $\Delta$while locally the uncertainty grows as
k grows (depending on $\xi$). A good choice of the shift $\xi$ seems
to be $\xi=\Delta$so that the Gaussian overlap increasing the uncertainty,
but are removed from each other by enough so that the local uncertainty
increases by a large amount. we end up with 

\[
\psi_{in}^{MD}(Q)=\frac{1}{\sqrt{{\displaystyle \sum_{i=0}^{m}\sum_{j=0}^{m}\pi\Delta^{2}e}xp[\frac{-(i-j)^{2}}{2}]}}\sum_{l=0}^{k}e^{-\frac{(Q_{1}+l\Delta)^{2}+(Q_{2}-l\Delta)^{2}}{2\Delta^{2}}}\]

\begin{figure}
\subfigure[]{\includegraphics[scale=0.4]{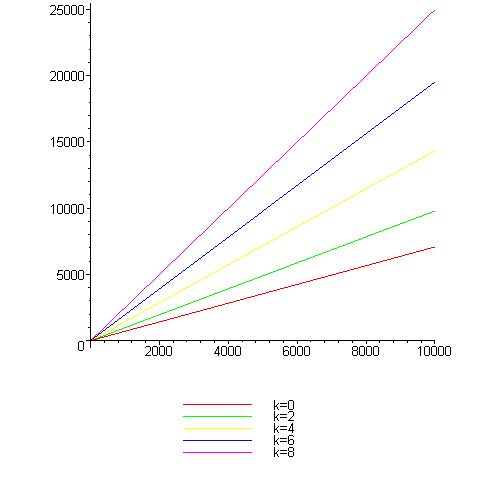}}\subfigure[$\delta=0.11;\epsilon=-0.05 \; \langle Q \rangle \rightarrow 21;$]{\includegraphics[scale=0.4]{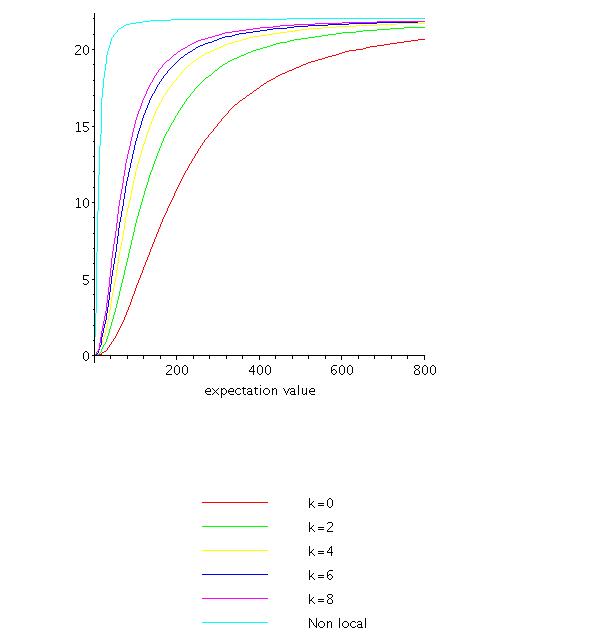}}

\caption{\label{newmethod}Comparing the local (red line), sum of Gaussian
(green,yellow blue and purple lines) and non local (light blue line)
methods for reaching the weak sum $(\sigma_{z}^{A}+\sigma_{z}^{B})_{w}=\frac{\left\langle \uparrow\downarrow+\downarrow\uparrow+\uparrow\uparrow+\downarrow\downarrow\right|\sigma_{z}^{A}+\sigma_{z}^{B}\left|\uparrow\downarrow(1+\epsilon)+\downarrow\uparrow(-1+\epsilon)+\delta\uparrow\uparrow\right\rangle }{\left\langle \uparrow\downarrow+\downarrow\uparrow+\uparrow\uparrow|\uparrow\downarrow(1+\epsilon)+\downarrow\uparrow(-1+\epsilon)+\delta\uparrow\uparrow\right\rangle }=\frac{2\delta}{2\epsilon+\delta}$}

The plots are the expectation values vs the uncertainty with $\delta=0.11$,
$\epsilon=-0.05$ we have $(\sigma_{z}^{A}+\sigma_{z}^{B})_{w}=22$ 

in (a) we see the local uncertainty as a function of different values
of k (for k=0 we have the standard local method). In (b) we can see
that as k grows the measurement approaches the non local limit.
\end{figure}

The local uncertainty for k=0 is just $\frac{\Delta}{\sqrt{2}}$ as
can be expected since this is just two local measuring devices. For
other values of k the uncertainty grows as can be seen in fig \ref{newmethod}.

\newpage{}

\section{Weak measurements of a non local product \label{sec:Joint-product}.}

Recently Resch and Steinberg \cite{ReschSteinberg} devised a method
for making measurements of the weak values of non local products.
This method, uses local weak measurements to extract what they call
{}``joint weak values'' . The method was further developed by Lundeen
and Resch \cite{ReschLundeen} with the title {}``practical measurements
joint weak values and their connection to the annihilation operator''.
It was later proposed by Mitchison, Jozsa and Popescu \cite{squential}
that such measurements be used for sequential weak measurements .
In this section we will investigate these methods and show that not
only do they depend on the local uncertainty, but that the uncertainty
in the final measurement is quadratic in $\Delta$ (the initial uncertainty
of each Gaussian) rather than  linear (as in the sum).

\subsection{Joint weak values\label{sub:Joint-weak-values.}.}

For a measuring device prepared in the state \begin{equation}
\psi_{AB-in}^{MD}(Q)=(\Delta^{2}\frac{\pi}{2})^{-1/2}e^{-(Q_{A}^{2}+Q_{B}^{2})/{4\Delta^{2}}}\label{md-inresch}\end{equation}
 the general pre and post selection (\ref{psi-in2},\ref{phi-fin2})
and the measurement Hamiltonian $H_{m}=g(t)(P_{A}A+P_{B}B)$ , we
have the following formulas for obtaining the weak values (after the
measurement).

\begin{equation}
Re(AB)_{w}=2\langle Q_{A}Q_{B}\rangle-Re(A_{w}^{*}B_{w})\label{Resch}\end{equation}
\begin{equation}
Im(AB)_{w}=\frac{4\Delta^{2}}{\hbar}\langle Q_{A}P_{B}\rangle-Im(A_{w}^{*}B_{w})\label{Reschim}\end{equation}

\begin{equation}
Re(AB)_{w}=\langle Q_{A}Q_{B}\rangle-\frac{4\Delta^{4}}{\hbar^{2}}\langle P_{A}P_{B}\rangle\label{laudeen}\end{equation}

\begin{equation}
Im(AB)_{w}=\frac{2\Delta^{2}}{\hbar}\left(\langle Q_{A}P_{B}\rangle-\langle P_{A}Q_{B}\rangle\right)\label{laudeenim}\end{equation}
These values depend on the correlations of the measuring devices and
on the local weak values and are derived by using a second order expansion
of the measuring device probability function $\psi_{AB-fi}^{MD}\psi_{AB-fi}^{MD*}$
with \begin{equation}
\psi_{AB-fi}^{MD}(Q)\approx\{1-A_{w}P_{A}-B_{w}P_{B}+\frac{1}{2}(A^{2})_{w}P_{A}P_{A}+\frac{1}{2}(B^{2})_{w}P_{B}P_{B}+\frac{1}{2}(AB)_{w}P_{A}P_{B}+O(\Delta^{-6})\}\psi_{AB-in}^{MD}(Q)\label{eq:MD-fin-second}\end{equation}
being the second order expansion of the final state of the measuring
device ($P\sim(\Delta^{-2})$). The measuring device is prepared in
such a way (\ref{md-inresch}) that all inner products of an odd number
of operators on the same particle ($\hat{Q}_{A};\hat{P}_{A}$ or $\hat{Q}_{B};\hat{P}_{B}$)
go to zero so that we get the results (\ref{Resch}-\ref{laudeenim}).

\begin{figure}
\subfigure[$\delta=0.11;\epsilon=-0.05 \; \langle Q \rangle \rightarrow 21$]{\includegraphics[scale=0.4]{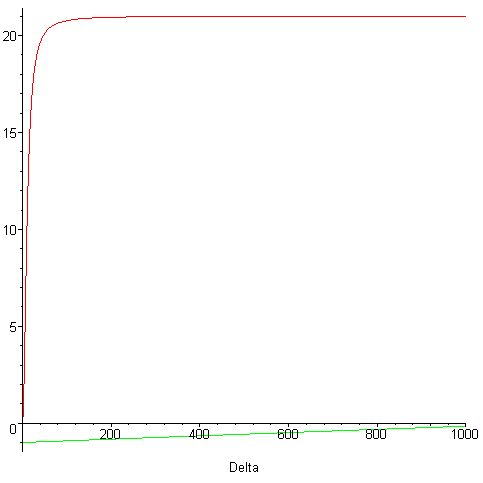}}\subfigure[$\delta=0.11;\epsilon=-0.05 \; \langle Q \rangle \rightarrow 21;$]{\includegraphics[scale=0.4]{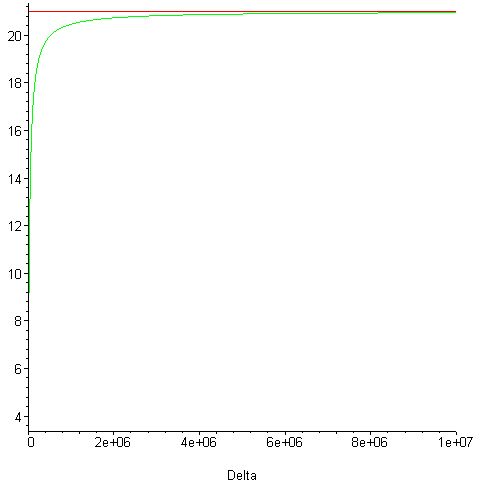}}

\caption{\label{poduct1}Comparing the local (green line) and non local {[}non
physical] (red line) methods for reaching the weak value of a product.
$(\sigma_{z}^{A}\sigma_{z}^{B})_{w}=\frac{\left\langle \uparrow\downarrow+\downarrow\uparrow+\uparrow\uparrow+\downarrow\downarrow\right|\sigma_{z}^{A}\sigma_{z}^{B}\left|\uparrow\downarrow(1+\epsilon)+\downarrow\uparrow(-1+\epsilon)+\delta\uparrow\uparrow\right\rangle }{\left\langle \uparrow\downarrow+\downarrow\uparrow+\uparrow\uparrow|\uparrow\downarrow(1+\epsilon)+\downarrow\uparrow(-1+\epsilon)+\delta\uparrow\uparrow\right\rangle }=\frac{\delta-2\epsilon}{2\epsilon+\delta}$}

The plots are the expectation values for different ranges of the uncertainty
($\Delta=0..1000;$$\Delta=0..10^{7})$. For $\delta=0.11$, $\epsilon=-0.05$
we have $(\sigma_{z}^{A}\sigma_{z}^{B})_{w}=21$ while the local weak
values are $(\sigma_{z}^{A})_{w}=211$, $(\sigma_{z}^{B})_{w}=-189$
(plots for the probability density can be seen in fig \ref{fig:locvnloc},\ref{fig:locvnloc2})

It can easily be seen that the non local method (red line) converges
much faster then the local method (green line). If we want to make
a precise measurements (deviation of 1\% and uncertainty of 10\%)
an ensemble of $n=1.5\times10^{12}$ is required for the non local
method while an ensemble of $n=2.5\times10^{3}$is required for the
non local method.
\end{figure}

Since for weak measurements the correlations for the two different
measuring devices are very weak, there is only a very slight shift
from zero in the different correlation functions. These slight shifts
are of second order (in $\Delta^{-2}$). But because the first order
expectation value is zero, they are the leading terms. As can be expected
the initial error for the measuring device is proportional to$\Delta^{2}$
rather than  $\Delta$. For the initial state we have:

\begin{equation}
\Delta_{QQ}=\Delta^{2}\label{errqq}\end{equation}
 \begin{equation}
\Delta_{PP}=\frac{1}{4\Delta^{2}}\label{errpp}\end{equation}
again the size of the ensemble ($n$) required for a precise measurement
depends on the overall measurement error. This means that the ensemble
for such joint weak measurements is much larger then the one required
for the measurement of a sum. Using the example with the states presented
previously (\ref{eq:expre},\ref{eq:expost}) \begin{equation}
(\sigma_{z}^{A}\sigma_{z}^{B})_{w}=\frac{\left\langle \uparrow\downarrow+\downarrow\uparrow+\uparrow\uparrow+\downarrow\downarrow\right|\sigma_{z}^{A}\sigma_{z}^{B}\left|\uparrow\downarrow(1+\epsilon)+\downarrow\uparrow(-1+\epsilon)+\delta\uparrow\uparrow\right\rangle }{\left\langle \uparrow\downarrow+\downarrow\uparrow+\uparrow\uparrow|\uparrow\downarrow(1+\epsilon)+\downarrow\uparrow(-1+\epsilon)+\delta\uparrow\uparrow\right\rangle }=\frac{\delta-2\epsilon}{2\epsilon+\delta}\label{eq:nlprod1}\end{equation}
and again taking $\delta=0.11$, $\epsilon=-0.05$ (fig \ref{poduct1})
we can see for a precise measurements (deviation of 1\% and uncertainty
of 10\%) an ensemble of $n=1.5\times10^{12}$ is required for the
joint weak measurements method. Since a method for making direct non
local weak measurements has not yet been found, we compare this method
with a non physical non-local measurement (involving non local interactions).
Such a method would require an ensemble of $n=2.5\times10^{3}$ for
a precise measurement.

As we have shown, this joint weak measurement method is an inefficient
measurement of the non local product. It is actually a lot worse then
the joint measurement of a non local sum (compared with the direct
non local weak measurement). 

\begin{figure}
\subfigure[$\delta=0.1;\epsilon=10\; \langle Q \rangle \rightarrow 1$]{\includegraphics[scale=0.4]{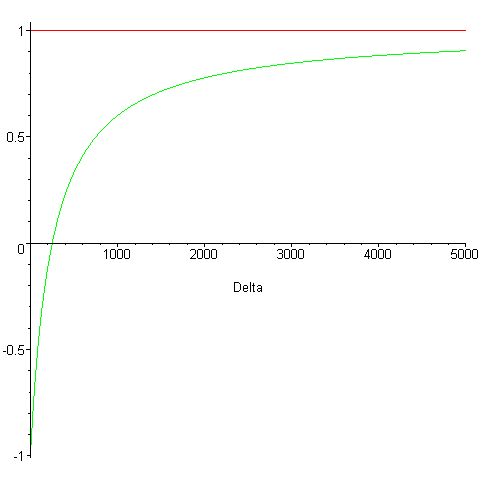}}\subfigure[$\delta=1;\epsilon=0.55 \; \langle Q \rangle \rightarrow -101;$]{\includegraphics[scale=0.4]{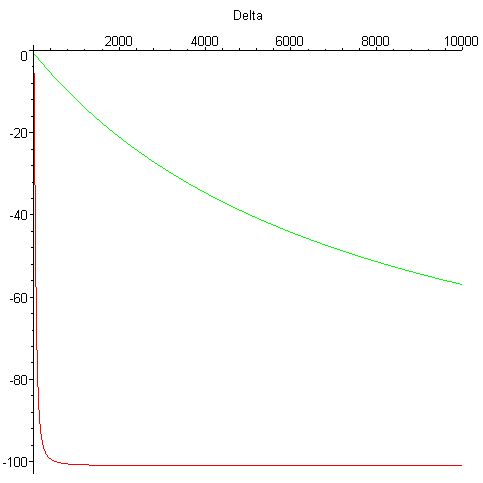}}

\subfigure[$\delta=-1;\epsilon=0.6\; \langle Q \rangle \rightarrow -11$]{\includegraphics[scale=0.4]{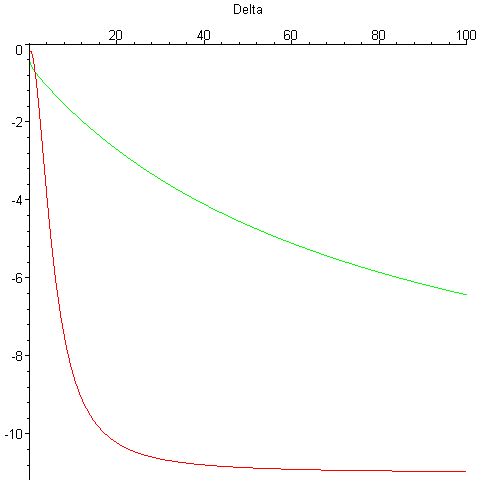}}\subfigure[$\delta=1;\epsilon=1 \; \langle Q \rangle \rightarrow -\frac{1}{3}$]{\includegraphics[scale=0.4]{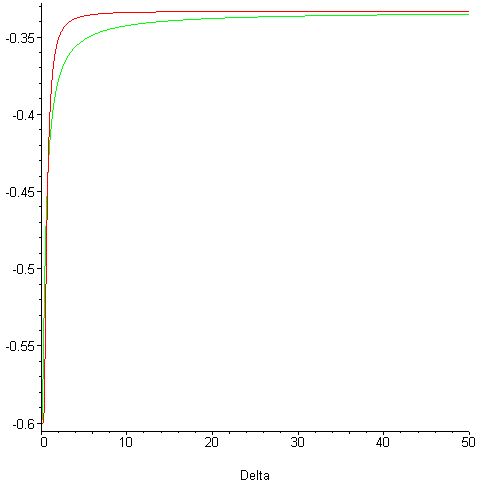}}

\caption{\label{poduct2}Comparing the local (green line) and non local {[}non
physical] (red line) methods for reaching the weak value of a product.$(\sigma_{z}^{A}\sigma_{z}^{B})_{w}=\frac{\left\langle \uparrow\downarrow+\downarrow\uparrow+\uparrow\uparrow+\downarrow\downarrow\right|\sigma_{z}^{A}\sigma_{z}^{B}\left|\uparrow\downarrow(1+\epsilon)+\downarrow\uparrow(-1+\epsilon)+\delta\uparrow\uparrow\right\rangle }{\left\langle \uparrow\downarrow+\downarrow\uparrow+\uparrow\uparrow|\uparrow\downarrow(1+\epsilon)+\downarrow\uparrow(-1+\epsilon)+\delta\uparrow\uparrow\right\rangle }=\frac{\delta-2\epsilon}{2\epsilon+\delta}$ }

The plots are for different values of $\delta$and $\epsilon$.
\end{figure}

Although there is no method for directly measuring the non local product
, the example above ( \ref{eq:nlprod1}) is a special case where we
can make a good measurement of the product by using a relation between
the sum, the modular sum and the product. For a modular sum and a
product (of two spin $1/2$ operators) we have the following relation:\begin{equation}
\sigma_{z}^{A}\sigma_{z}^{B}=(\sigma_{z}^{A}+\sigma_{z}^{B})\; mod\;4-1\label{modprod}\end{equation}
using this relation we can calculate the product from the modular
sum. This is not much help since we don't know how to measure the
modular sum directly, but since in our example there is zero probability
for the $\downarrow\downarrow$state, the sum is the same as the modular
sum. Thus by measuring the sum we can actually get the value of the
product. We already know how to measure the sum so we can use this
method for making a good measurement of the weak value of the product.
This method works for any two spin $1/2$ system, as long as one of
the 4 possible local product states $\uparrow\uparrow$; $\downarrow\downarrow$;
$\uparrow\downarrow$;$\downarrow\uparrow$ has zero probability.
(although we need to look at $\sigma_{z}^{A}-\sigma_{z}^{B}$ if its
one of the spin zero states that has zero probability).

It is important to point out another practical consideration for the
{}``joint weak measurements''. As we have already shown, the pointer
is set in such a way that it points at zero both before and after
the interaction. The weak value is reached only by looking at second
order effects. In a practical situation there might be a problem getting
the measuring device set precisely on zero. Looking at (\ref{eq:MD-fin-second})
we can see that in the case where some of the first order terms are
nonzero (after integration) we get contributions from $A_{w};B_{w};(A^{2})_{w};(B^{2})_{w}$.
For the standard weak measurements, Jozsa \cite{complexweak} showed
that when using an arbitrary (weak) measuring device, the deviation
from the weak value in the final result depends on properties of the
measuring device and the weak value alone. This dependence on other
weak values makes it even harder to make practical measurements of
join weak values.

\subsection{Sequential weak measurements.}

Weak values of a product were shown to be of interest by Mitchison,
Jozsa and Popescu \cite{squential} when used for measuring the weak
value of the product of the operators at different times, $(B_{t_{2}}A_{t_{1}})_{w}=\frac{\langle\Phi|B{}_{t_{2}}VA_{t_{1}}|\Psi\rangle}{\langle\Phi|\Psi\rangle}$
where $V$ is the unitary evolution of the system between $t_{1}$and
$t_{2}$. The term used for the result is sequential weak values and
it was shown that in a double interferometer experiment where the
sequential strong measurement cannot be made because one measurement
(at $t_{1}$) effects the other (at $t_{2}$), the sequential weak
values can still be calculated. In that paper two methods were shown
for calculating the sequential weak value of the number of photons
going through an double interferometer. The first was the method of
joint weak measurements which as we have shown cannot be considered
a good measurement method. The second method involves three local
weak measurements of different projection operators and some prior
knowledge of the system. In that context it is interesting to discuss
the meaning of the sequential weak value. Although the sequential
weak value has a definite value, it cannot be measured directly and
therefore cannot be considered an element of reality (as defined by
Vaidman\cite{elements} ) like standard weak measurements.

\subsection{Weak measurements on ensembles of random non local systems with the
same weak value.}

The concept of weak measurements on random systems with the same weak
value mentioned in sec \ref{sub:WMrand} can also be used to examine
joint weak values. Since both the values $\langle Q_{A}Q_{B}\rangle$
and $\langle P_{A}P_{B}\rangle$ depend on $Re(A_{w}^{*}B_{w})$.
A set of random states with the same value for $(AB)_{w}$ but different
values for $(A_{w}^{*}B_{w})$ will give us very different values
for $\langle Q_{A}Q_{B}\rangle$ and $\langle P_{A}P_{B}\rangle$.
Since both cannot be measured at the same time, formula \ref{laudeen}
cannot be used for calculating the weak value. Let us look at the
following example: We start with an ensemble of systems pre selected
in any of the states \begin{equation}
|\Psi_{i}\rangle_{AB}=\frac{1}{\sqrt{2\epsilon_{i}^{2}+\delta_{i}^{2}+2}}\left|\uparrow\downarrow(1+\epsilon_{i})+\downarrow\uparrow(-1+\epsilon_{i})+\delta_{i}\uparrow\uparrow\right\rangle \label{eq:expre2}\end{equation}
and all post selected in the same state\begin{equation}
|\Phi_{i}\rangle_{AB}=\frac{1}{2}|\uparrow\downarrow+\downarrow\uparrow+\uparrow\uparrow+\downarrow\downarrow\rangle\label{eq:expost2}\end{equation}
With that the states $|\Psi_{i}\rangle$chosen with $Re(\epsilon)=0$,
$Im(\epsilon)=\frac{-Im(\delta)}{2}$ and $Re(\delta)=3Im(\delta)$so
that \begin{equation}
(\sigma_{z}^{A}\sigma_{z}^{B})_{w}=\frac{\delta-2\epsilon}{2\epsilon+\delta}=1+\frac{2}{3}i\end{equation}
 defining $\delta''\equiv Im(\delta)$ \begin{equation}
(\sigma_{z}^{A})_{w}=\frac{\delta+2}{2\epsilon+\delta}=1+\frac{2}{3}\delta''+\frac{i}{3}\end{equation}
\begin{equation}
(\sigma_{z}^{B})_{w}=\frac{\delta+2}{2\epsilon+\delta}=1-\frac{2}{3}\delta''+\frac{i}{3}\end{equation}
 so that \[
(\sigma_{z}^{A})_{w}^{*}(\sigma_{z}^{B})_{w}=\frac{\delta+2}{2\epsilon+\delta}=\frac{4}{3}+\frac{4\delta''^{2}}{9}+i\frac{4}{9}\delta''\]

Using (\ref{Resch}-\ref{laudeenim}) we have the following values
for the joint measurements of the measuring devices.\begin{equation}
\langle Q_{A}Q_{B}\rangle=\frac{7}{6}+\frac{2\delta''^{2}}{9}\end{equation}
\begin{equation}
\frac{4\Delta^{4}}{\hbar^{2}}\langle P_{A}P_{B}\rangle=\frac{1}{6}+\frac{2\delta''^{2}}{9}\end{equation}

\begin{equation}
\frac{2\Delta^{2}}{\hbar}\langle Q_{A}P_{B}\rangle=\frac{2}{6}+\frac{2}{9}\delta\end{equation}
\begin{equation}
\frac{2\Delta^{2}}{\hbar}\langle P_{A}Q_{B}\rangle=-\frac{2}{6}+\frac{2}{9}\delta\end{equation}
Since it is not possible to look at all these observables at the same
time, their correlations will be lost if the spread $\delta''$ is
large enough. 

If we allow ourselves to look at the measuring device in any (local)
way, we can measure the value \begin{equation}
\langle(Q_{A}-\frac{2\Delta^{2}}{\hbar}P_{A})(Q_{B}+\frac{2\Delta^{2}}{\hbar}P_{B})\rangle=Re(AB)_{w}+Im(AB)_{w}\end{equation}
 which will give us the right result for $Im(AB)_{w}=0$ . In our
case this will not be enough. As we already saw at the end of section
\ref{sub:Joint-weak-values.}, our example is one where we can measure
the product by measuring the sum. Using this method we can make the
measurement of the weak value for our random ensemble.

\newpage{}

\section{The meaning of non local weak values \label{sec:The-meaning-of}.}

While eigenvalues, the outcomes of ideal strong measurements, and
expectation values, the average of reading of ideal measurements over
an ensemble of identically prepared systems are the basic textbook
concepts of quantum mechanics, weak values have less solid foundations.
We can still hear the echoes of the controversy of the days when weak
values were introduced \cite{Replyaharonov,Replyleggett,Replyperes}.
The justification of considering weak values as a description of a
pre- and post-selected quantum system relies on the universality of
influence of the coupling to a variable in the limit of its weakness.
The pointer variable prepared in a natural way (see Jozsa for some
limitations \cite{complexweak}) shifts due to weak measurement coupling
as if it were coupled to a classical variable with the value equal
to the weak value. (Note also that weak measurement performed on pre-selected
ensemble show the expectation value even though we do not find the
eigenvalues in the process of measurement.) Recently a method for
making indirect measurements of weak variables was introduced \cite{Johansen}.
This method together with the indirect methods mentioned in the last
two sections remindes us that the weak value is the result of a calculation
involving the pre selection, post selection and an hermitian operator. 

We have shown that weak value have the following property: Although
an ensemble is required to measure the weak value with any accuracy,
this ensemble need not be made of identical systems. The only requirement
is that the weak value remain the same for all systems. This is a
property of weak measurements which is related to the effective coupling
between the measuring device and the weak value. It shows that weak
measurements are affected by the weak value alone and not by the other
properties of the pre, and post selected states. Again we see that
they are a property of the sytem that can be measured.

In this light we can see that those weak values which cannot be measured
using a direct weak measurement have a lesser status. There is no
method for creating an effective coupling between the measuring device
and weak value of a product$(AB)_{W}$. Again we see that indirect
measurements of non local weak values have a lesser status than standard
weak measurements.\newpage{}

\section{Conclusions}

We discussed some methods for measuring the weak values of non local
observables such as the sum of two observables (belonging to different
particles) and the product of two observables (belonging to different
particles). For the case of a non local sum we found a method for
making non local weak measurements using an entangled measuring device.
For the case of a product no such method has been found.

Indirect methods for reaching the non local weak value were analyzed
and compared to direct methods. The size of the ensemble required
to make a precise measurement was used to distinguish between efficient
and inefficient methods. It was shown that for indirect methods the
size of the ensemble required is not related directly to the uncertainty
in the variable to be measured. These indirect methods are therefore
inefficient.

When measuring a sum there are special cases where local measurements
are almost as efficient as non local measurements, these are cases
where the non local uncertainty is the same as the local uncertainty
( \ref{weakuncer}). For a small uncertainty in the non local variable
and a large uncertainty in the local variable we showed (fig \ref{fig:locvnloc}-\ref{fig:locvnlocexp})
that the non local method is much better then the local method.

The method of {}``joint weak values'' for measureing the product
is even more inefficient since even in the case where both the local
and the non local uncertainty are small , local joint weak measurements
require a large ensemble. This is because the deviation from zero
in the observables $Q_{A}Q_{B}$, $P_{A}P_{B}$, $Q_{A}P_{B}$and
$P_{A}Q_{B}$ is very small compared with the width (the uncertainty)
of these observables. For that reason such methods seem very impractical.
However for some specific cases, we found a way for measuring the
weak value of the product using the method described for the weak
measurement of a sum. 

The interpretation of weak values as {}``elements of reality'' \cite{elements}
depends on the  measuring device pointer being effectively coupled
to the weak value. This effective coupling between the measuring device
and the weak value allows us to measure the weak value using an ensamble
of random systems with the same weak value. The above does not hold
for joint weak measurements. It is therefore not clear if weak values
of a non local product can be thought of as elements of reality. 

Although we can write the expression for any weak value, for some
non local weak values (such as a product) there is no efficient measurement
procedure. It follows that those weak values which cannot be measured
using direct weak measurements have a lesser status then those weak
values that can be measured directly.\\
\\
This work has been supported in part by the European Commission under
the Integrated Project Qubit Applications (QAP) funded by the IST
directorate as Contract Number 015848 and by grant 990/06 of the Israel
Science Foundation. 

\bibliographystyle{unsrt} 

\newpage{}

\bibliographystyle{amsplain}

\begin{thebibliography}{10}

\bibitem{Landau}
L.~Landau and R.~Peierls.
\newblock Erweiterung des unbestimmtheitsprinzips f?r die relativistische
  quantentheorie.
\newblock {\em Z. Phys. 69, 56}.

\bibitem{InstanNL}
Lev Vaidman.
\newblock Instantaneous measurement of nonlocal variables.
\newblock {\em Phys. Rev. Lett.}, 90(1):010402, Jan 2003.

\bibitem{QuantumMeasurmentbook}
Farid Ya~Khalili Vladimir B.~Braginsky.
\newblock Quantum measurement, 1995.

\bibitem{AAVnonlocal}
Yakir Aharonov, David~Z. Albert, and Lev Vaidman.
\newblock Measurement process in relativistic quantum theory.
\newblock {\em Phys. Rev. D}, 34(6):1805--1813, Sep 1986.

\bibitem{AharonovAlbert}
Yakir Aharonov and David~Z. Albert.
\newblock Can we make sense out of the measurement process in relativistic
  quantum mechanics?
\newblock {\em Phys. Rev. D}, 24(2):359--370, Jul 1981.

\bibitem{vonNeumann}
John von Neumann.
\newblock {\em Mathematische Grundlagen der Quantenmechanik}.
\newblock 1932.
\newblock Reprinted: Dover Publications (1943); Presses Universitaires de
  France (1947); Madrid, Institute de Matematicas ``Jorge Juan'' (1949).
  Translated from German by Robert T. Beyer, Princeton University Press (1955).

\bibitem{spin100}
Yakir Aharonov, David~Z. Albert, and Lev Vaidman.
\newblock How the result of a measurement of a component of the spin of a
  spin-1/2 particle can turn out to be 100.
\newblock {\em Phys. Rev. Lett.}, 60(14):1351--1354, Apr 1988.

\bibitem{timeinQM}
Yakir Aharonov and Lev Vaidman.
\newblock The two-state vector formalism of qauntum mechanics, 2002.

\bibitem{ABL}
Yakir Aharonov, Peter~G. Bergmann, and Joel~L. Lebowitz.
\newblock Time symmetry in the quantum process of measurement.
\newblock {\em Phys. Rev.}, 134(6B):B1410--B1416, Jun 1964.

\bibitem{ReschSteinberg}
K.~J. Resch and A.~M. Steinberg.
\newblock Extracting joint weak values with local, single-particle
  measurements.
\newblock {\em Phys. Rev. Lett.}, 92(13):130402, Mar 2004.

\bibitem{Resch2}
K.~J. Resch.
\newblock Practical weak measurement of multiparticle observables.
\newblock {\em J. Opt. B: Quantum Semiclass. Opt.}, 6:482--487, 2004.

\bibitem{ReschLundeen}
J.~S. {Lundeen} and K.~J. {Resch}.
\newblock Practical measurement of joint weak values and their connection to
  the annihilation operator.
\newblock {\em Physics Letters A}, 334:337--344, January 2005.

\bibitem{squential}
G.~{Mitchison}, R.~{Jozsa}, and S.~{Popescu}.
\newblock {Sequential weak measurement}.
\newblock {\em Pys Rev A}, 76(6):062105--+, December 2007.

\bibitem{properties}
Yakir Aharonov and Lev Vaidman.
\newblock Properties of a quantum system during the time interval between two
  measurements.
\newblock {\em Phys. Rev. A}, 41(1):11--20, Jan 1990.

\bibitem{Weakexp1}
G.~J. {Pryde}, J.~L. {O'Brien}, A.~G. {White}, T.~C. {Ralph}, and H.~M.
  {Wiseman}.
\newblock {Measurement of Quantum Weak Values of Photon Polarization}.
\newblock {\em Physical Review Letters}, 94(22):220405--+, June 2005.

\bibitem{weakexp2}
N.~W.~M. {Ritchie}, J.~G. {Story}, and R.~G. {Hulet}.
\newblock {Realization of a measurement of a ``weak value''}.
\newblock {\em Physical Review Letters}, 66:1107--1110, March 1991.

\bibitem{elements}
L.~{Vaidman}.
\newblock {Weak-measurement elements of reality}.
\newblock {\em Foundations of Physics}, 26:895--906, July 1996.

\bibitem{complexweak}
R.~{Jozsa}.
\newblock {Complex weak values in quantum measurement}.
\newblock {\em Phys Rev A}, 76(4):044103--+, October 2007.

\bibitem{nonlocaltimesymmetric}
L.~{Vaidman} and I.~{Nevo}.
\newblock {Nonlocal Measurements in the Time-Symmetric Quantum Mechanics}.
\newblock {\em International Journal of Modern Physics B}, 20:1528--1535, 2006.

\bibitem{Replyaharonov}
Y.~Aharonov and L.~Vaidman.
\newblock Aharonov and vaidman reply.
\newblock {\em Phys. Rev. Lett.}, 62(19):2327, May 1989.

\bibitem{Replyleggett}
A.~J. Leggett.
\newblock Comment on \char96{}\char96{}how the result of a measurement of a
  component of the spin of a spin-(1/2 particle can turn out to be
  100\char39{}\char39{}.
\newblock {\em Phys. Rev. Lett.}, 62(19):2325, May 1989.

\bibitem{Replyperes}
Asher Peres.
\newblock Quantum measurements with postselection.
\newblock {\em Phys. Rev. Lett.}, 62(19):2326, May 1989.

\bibitem{Johansen}
L.~M. {Johansen}.
\newblock {Reconstructing weak values without weak measurements}.
\newblock {\em Physics Letters A}, 366:374--376, July 2007.

\bibitem{Kedem2010}
 {Kedem, Yaron and Vaidman, Lev},
\newblock{Modular values and weak values of quantum observables},
\newblock {\em Physical review letters} 105, 23041, 2010.

\bibitem{Brodutch2009}
 {Brodutch, Aharon and Vaidman, Lev},
  \newblock{Measurements of non local weak values},
\newblock {\em Journal of Physics: Conference Series} 174 012004 2009




\end{thebibliography}

\end{document}